\documentclass[9pt,twocolumn,twoside]{opti_preprint}

\setboolean{shortarticle}{false}

\usepackage{soul}

\title{Control of the temporal and polarization response of a multimode fiber}

\author[1,*]{Mickael Mounaix}
\author[1]{Joel Carpenter}

\affil[1]{School of Information Technology and Electrical Engineering, The University of Queensland, Brisbane QLD 4072, Australia}

\affil[*]{Corresponding author: m.mounaix@uq.edu.au}

\begin{abstract}
Control of the spatial and temporal properties of light propagating in disordered media have been demonstrated over the last decade using spatial light modulators. Most of the previous studies demonstrated spatial focusing to the speckle grain size, and manipulation of the temporal properties of the achieved focus. In this work, we demonstrate temporal control of the total impulse response integrated over all the spatial and polarization modes propagating through a multimode fiber. We notably demonstrate a global enhancement of light intensity at a chosen arrival time, as well as attenuating light intensity at an arbitrary delay. We also demonstrate the full polarization control of such engineered states and a multiple control at different delay times, which opens interesting perspectives for non-linear imaging through complex systems and high power fiber lasers.  
\end{abstract}

\begin{document}

\maketitle
\pagestyle{plain}
\ifthenelse{\boolean{shortarticle}}{\abscontent}{}

\section{Introduction}

Light propagation in disordered systems has been extensively studied over the past 40 years~\cite{goodman_fundamental_1976}. Transmitted light through such materials takes the form of a speckle pattern, that is the signature of the interference between a huge number of modes within the sample. Although this speckle looks random, this mixing of light is nonetheless linear and thus deterministic. Wavefront shaping has revolutionized the spatial control of coherent light beams thanks to spatial light modulators (SLM) in such disordered systems~\cite{rotter_light_2017,mosk_controlling_2012} , such as biological tissue, white paint or multimode fibers. Different approaches have been developed to control the propagation of coherent light that has suffered from scattering, such as iterative optimization~\cite{vellekoop_focusing_2007,vellekoop_feedback-based_2015}, optical phase conjugation~\cite{yaqoob_optical_2008} and the measurement of the optical transmission matrix~\cite{popoff_measuring_2010}. These methods have been further extended to the control of polarization~\cite{guan_polarization_2012}, spectral~\cite{andreoli_deterministic_2015, paudel_focusing_2013} and temporal properties of light~\cite{mounaix_deterministic_2016}, and widely used for imaging~\cite{cizmar_exploiting_2012,choi_scanner-free_2012, ploschner_seeing_2015, boonzajer_flaes_robustness_2018} and optical manipulation~\cite{leite_three-dimensional_2018}.

The spatial and temporal distortions of scattered light are coupled in disordered systems. Therefore, the temporal properties of a spatio-temporal speckle field can be adjusted with spatial-only control. In previous works, manipulating the temporal properties of scattered light was achieved with spatio-temporal focusing with a single SLM. The output pulse is focused on a single speckle grain, and the spectral properties of the focus are controlled to ensure a short duration of the output pulse. Spatio-temporal focusing of the output pulse can be reached for instance with optimization algorithms~\cite{aulbach_control_2011,katz_focusing_2011}, digital phase conjugation~\cite{morales-delgado_delivery_2015}, pulse shaping methods~\cite{mccabe_spatio-temporal_2011}, or via the knowledge of the transmission matrix of the sample~\cite{mounaix_spatiotemporal_2016, mounaix_deterministic_2016}. However, the output pulse has only a short duration at this specified focus position, while the background speckle remains temporally elongated~\cite{mounaix_temporal_2017}. The temporal control is then limited to a specific spatial mode (speckle position or a specific spatial pattern), at the expense of undefined temporal behavior for the other spatial modes. A control of the full speckle pattern over a certain spectral bandwidth has been proposed based on the transmission matrix, with the time delay operator~\cite{carpenter_observation_2015,xiong_spatiotemporal_2016,ambichl_super-_2017-1}. Nonetheless, the limited controlled spectral bandwidth does not allow a full temporal control of the spatio-temporal speckle.

\begin{figure*}[htbp]
\centering
\includegraphics[width=\linewidth]{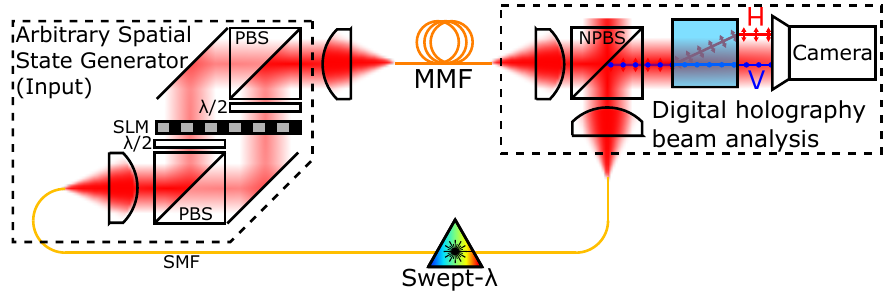}
\caption{Apparatus of the experimental setup. SMF: single mode fiber; SLM: spatial light modulator; PBS: polarizing beam splitter; MMF: multimode fiber; NPBS: non-polarizing beam splitter}
\label{fig:exp_setup}
\end{figure*}

In this paper, we propose a novel method to adjust the temporal properties of the total impulse response integrated over all the spatial and polarization modes propagating through a multimode fiber. After experimentally measuring the Multi-Spectral Transmission Matrix (MSTM) of the fiber, we calculate the Time Resolved Transmission Matrix (TRTM) with a Fourier Transform. Exploiting the TRTM, we demonstrate both temporal enhancement and attenuation at specific delay times at the expense of undefined behaviour in the spatial domain. We also demonstrate control of the often neglected polarization degree of freedom, which represents half of all the modes supported.

\section{Experimental setup and measurement of the Multi Spectral Transmission Matrix of a multimode fiber}

At a given wavelength $\lambda$, the transmission matrix $U$ of a disordered system, such as a multimode fiber, linearly links the input field (adjustable with a SLM) to the output field (detectable with a camera). $U$ is a $N_{\text{input}}~\times~N_{\text{output}}$ matrix, with $N_{\text{input}}$ and $N_{\text{output}}$ the number of spatial and polarization modes at the input and at the output of the system. The transmission matrix has been widely used to manipulate light that has suffered from scattering or mode mixing, for focusing~\cite{popoff_measuring_2010} or imaging purposes~\cite{popoff_image_2010}. Recently, the transmission matrix has been extended to the spectral domain. The Multi Spectral Transmission Matrix~\cite{andreoli_deterministic_2015,mounaix_spatiotemporal_2016}, also known as the optical transfer function~\cite{carpenter_complete_2016}, is a stack of transmission matrices $U(\lambda)$ for a set of different input wavelength, that can be measured by sweeping the wavelength or using hyper-spectral imaging~\cite{boniface_rapid_2019}. The MSTM enables a spectral control of the output field exploiting the spectral diversity of the medium, and could be used for focusing a pattern at a given wavelength~\cite{andreoli_deterministic_2015,carpenter_complete_2016}, pulse shaping the output focus~\cite{mounaix_spatiotemporal_2016} and imaging through scattering samples~\cite{mounaix_transmission_2018}. 

We exploit the setup presented in Fig.~\ref{fig:exp_setup} to measure the Multi Spectral Transmission Matrix of a multimode fiber (MMF). The light source is a tunable continuous wave (CW) laser (New focus TLM-8700) operating between 1510~nm and 1620~nm. We use a phase-only spatial light modulator (Meadowlark P1920) to generate any spatial and polarization input state in amplitude and phase~\cite{carpenter_110x110_2014}. The SLM is located in the Fourier plane of a multimode fiber. The fiber is a 1m length of a 50~$\mu$m core radius step-index multimode fiber (Thorlabs M42L01), which theoritically supports approximately 127 spatial modes per polarization state~\cite{snyder_optical_2012}. The output field is measured with digital off-axis holography, using an external path-length matched reference beam and an InGaAs camera (Xenics Xeva-1.7-320). The two polarization states are measured on the left and the right side of the camera using a polarizer beam displacer. 

We use a swept-wavelength interferometer (SWI) to measure the multi spectral transmission matrix of the MMF~\cite{carpenter_110x110_2014,carpenter_complete_2016}. We generate all the spatial and polarization modes (a total of $N_{\text{input}}=254$~modes) on the input facet of the MMF using the SLM. For each individual mode, we sweep the laser and record the output field for all the wavelengths with the digital holography beam analysis. The camera is triggered at 10.6~GHz through the sweep with an external Mach-Zehnder interferometer that acts as a k-clock. Each output field is then digitally decomposed in the Laguerre-Gaussian basis ($N_{\text{output}}=650$~modes). The spectral resolution of the SWI enables the measurement of $N_{\lambda}=1273$ monochromatic transmission matrices, at a speed of 16 seconds per input mode. Additional details on the experimental setup are provided in the Supplemental Material.

\section{Enhancing the intensity over all spatial and polarization modes at an arbitrary delay time}

\begin{figure*}[htbp]
\centering
\includegraphics[width=\linewidth]{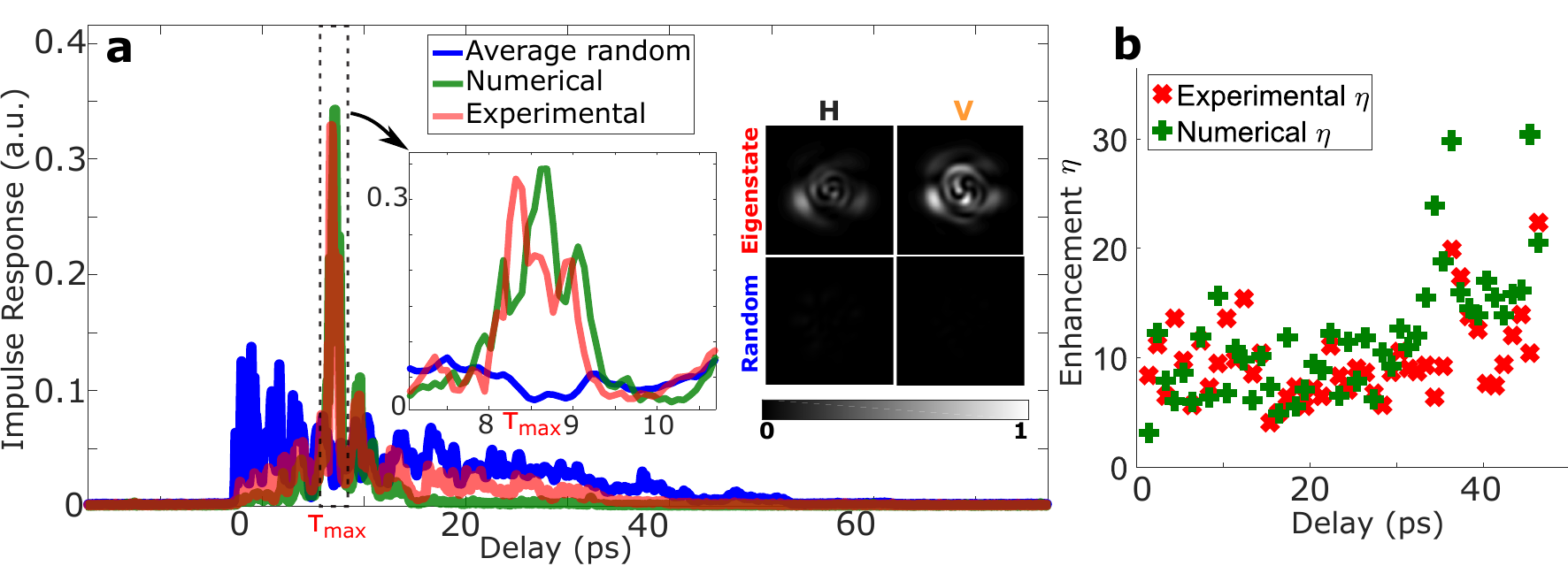}
\caption{Enhancing light intensity at a chosen delay time over all the spatial and polarization modes of a multimode fiber. (a) Enhancing light intensity at $\tau_{\text{max}}=8.7$~ps. Blue: Time of flight distribution of the multimode fiber. Red: experimental impulse response of the eigenstate maximizing the light intensity at $\tau_{\text{max}}$. Green: Numerical propagation of this eigenstate with the experimentally measured MSTM. Inset: zoom around $\tau_{\text{max}}$ ; Reconstructed output intensity at $\tau_{\text{max}}$ for the two polarization states for the eigenstate (top row) and a random combination of modes (bottom row).  (b) Enhancement ratio $\eta$ for different delay times. Red: enhancement ratios calculated with the experimentally measured impulse responses. Green: enhancement ratios calculated with the numerically propagated impulse responses.}
\label{fig:enhancing_impulse_response}
\end{figure*}

With a sufficient spectral sampling, the MSTM can be Fourier transformed to access the transmission matrices as a function of delay $U(\tau)$. The obtained Time Resolved Transmission matrix (TRTM) contains the full relationship between the spatiotemporal input field to the spatiotemporal output field. The TRTM has been exploited so far to focus light in space and time~\cite{mounaix_deterministic_2016} and to study spatio-temporal correlations~\cite{xiong_spatio-temporal_2018}. We show in the Supplemental Material a set of transmission matrices at different delay times of the multimode fiber, that we calculated from our measured MSTM. 

We demonstrate here how to exploit the TRTM to control the light intensity integrated over all spatial and polarization modes at an arbitrary delay $\tau_s$. The transmission matrix of the MMF at $\tau_s$, that we write $U_{\tau_s}$, is extracted from the TRTM. Initially developed in acoustics~\cite{prada_eigenmodes_1994,prada_decomposition_1996}, the Time Reversal Operator $U^{\dagger} U$, where the $\dagger$ operator stands for the conjugate transpose, has been extensively used in the monochromatic regime for selective focusing~\cite{popoff_exploiting_2011}, and the study of open channels~\cite{choi_transmission_2011,davy_transmission_2015,davy_universal_2015, yilmaz_transverse_2019}. Indeed, its eigenvalues are directly related to the total transmitted energy at the output of the disordered system~\cite{kim_maximal_2012}. In the time domain, the eigenvectors of $U_{\tau_s}^{\dagger} U_{\tau_s}$ have been used to study optimal transmission through scattering systems~\cite{hsu_broadband_2015}. For instance, in order to maximize the transmitted energy at the delay time $\tau_s$, the input state $E_{\text{input}}^{\text{max}}$ is the eigenvector of $U_{\tau_s}^{\dagger} U_{\tau_s}$ associated to the highest eigenvalue~\cite{xiong_spatio-temporal_2018}. Our experimental setup enables to measure all spatial modes in both polarizations at both ends of the fiber within $U(\lambda)$, and thus within $U_{\tau_s}$. Therefore launching $E_{\text{input}}^{\text{max}}$ is thus enhancing both the horizontal and the vertical polarization states at the output of the MMF at the delay time $\tau_s$. In contrast with digital phase conjugation of $U_{\tau_s}$~\cite{mounaix_deterministic_2016}, this eigenstate is not focusing light in a certain spatial mode or in a spatial position: it is enhancing the light intensity at $\tau_s$ for all the spatial and polarization modes. 

We experimentally demonstrate this enhancement of light intensity at an arbitrary delay time in Fig.~\ref{fig:enhancing_impulse_response}. The delay time $\tau=0$~ps corresponds to the arrival time of the first principal mode of the MMF~\cite{carpenter_observation_2015}. As an illustrative example, we maximize the transmitted intensity at $\tau_{\text{max}}=$~8.5~ps in Fig.~\ref{fig:enhancing_impulse_response}a. Once we display the eigenvector associated to the maximum eigenvalue of $U_{\tau_{\text{max}}}^{\dagger} U_{\tau_{\text{max}}}$ onto the SLM, we measure the impulse response $I(\tau)$ of the output state. For this purpose, we measure the spectrally resolved output field $E_{\text{output}}(\lambda)$ with the SWI, while the fixed input state is being displayed on the SLM. The impulse response is calculated from the superposition over all the $N_{\text{output}}$ output modes:
\begin{equation} \label{eq:impulse_response}
I(\tau)=\sum_{j=1}^{N_{\text{output}}} |\hat{E}_{\text{output}}(\tau,j)|^2
\end{equation}
with $\hat{E}_{\text{output}}(\tau,j)$ the Fourier transform of $E_{\text{output}}(\lambda)$ decomposed on the j-th output mode. The impulse response is then normalized. The calculated impulse response with Eq.~\ref{eq:impulse_response} is then equivalent to what would be measured by a large photodetector at the end of the fiber.

 \begin{figure*}[htbp]
\centering
\includegraphics[width=\linewidth]{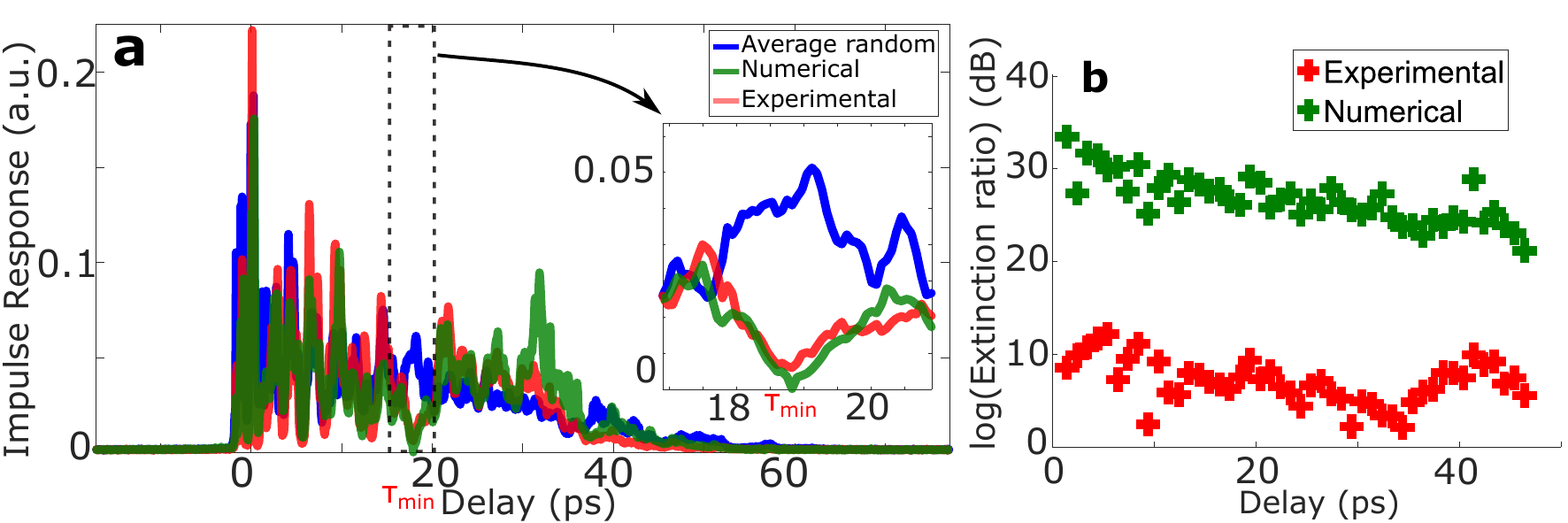}
\caption{Attenuating light intensity at a chosen delay time. (a) Attenuating light intensity at $\tau_{\text{min}}=19$~ps. Blue: Time of flight distribution of the multimode fiber. Red: experimental impulse response of the eigenstate minimizing the light intensity at $\tau_{\text{min}}$. Green: Numerical propagation of this eigenstate with the experimentally measured MSTM. Inset: zoom around $\tau_{\text{min}}$ (b) Attenuation ratio $\rho$ for different delay times in logarithm scale. Red: enhancement ratios calculated with the experimentally measured impulse responses. Green: enhancement ratios calculated with the numerically propagated impulse responses.}
\label{fig:extinction_impulse_response}
\end{figure*}

In Fig.~\ref{fig:enhancing_impulse_response}a., we plot the experimentally measured time of flight distribution of the multimode fiber in blue color as a reference impulse response. It corresponds to the average impulse response $I_{\text{rand}}$ over a set of $N_{\text{rand}}$~=~20 random combinations over all spatial and polarization input states. The experimentally measured impulse response $I_{\tau_{\text{max}}}$ of the maximum eigenstate at $\tau_{\text{max}}$ is plotted in red in Fig.~\ref{fig:enhancing_impulse_response}a, which clearly shows a sharp peak at $\tau_{\text{max}}$. This impulse response could be numerically predicted with the MSTM. For every wavelength $\lambda$, the expected output field $E_{\text{output}}^{\text{numerical}}(\lambda)$ reads $E_{\text{output}}^{\text{numerical}}(\lambda)~=~U(\lambda)~E_{\text{input}}^{\text{max}}$. From the set of $E_{\text{output}}^{\text{numerical}}(\lambda)$, we can use Eq.~\ref{eq:impulse_response} to calculate the expected impulse response $I_{\tau_{\text{max}}}^{\text{numerical}}$. We plot $I_{\tau_{\text{max}}}^{\text{numerical}}$ in green in Fig.~\ref{fig:enhancing_impulse_response}a, which is, as expected, very similar to the experimentally measured impulse response.

From the spectrally-resolved field measurements, we can reconstruct the output intensity at the delay time $\tau_{\text{max}}$ with a Fourier transform. On the inset of Fig.~\ref{fig:enhancing_impulse_response}a., we show the reconstructed intensity at $\tau_{\text{max}}$ for both polarization states for the enhancing eigenstate (top row). The output intensity is here spread over a combination of spatial modes for both the horizontal and the vertical polarization. As a contrast with digital optical phase conjugation of the TRTM~\cite{mounaix_deterministic_2016}, the output intensity is not spatially focused to a mode or to a speckle grain. As a comparison, we also show the reconstructed intensity of a random combination of input modes of both polarization at $\tau_{\text{max}}$, with the same color-bar (bottom row). 

Similarly to spatial focusing experiments~\cite{vellekoop_focusing_2007}, we define an enhancement ratio $\eta(\tau_{\text{max}})$ when enhancing the output intensity at delay time $\tau_{\text{max}}$. Based on the impulse responses plotted in Fig.~\ref{fig:enhancing_impulse_response}a., $\eta(\tau_{\text{max}})$ is defined as follows:
\begin{equation} \label{eq:enhancement_ratio}
\eta(\tau_{\text{max}}) = \frac{I_{\tau_{\text{max}}}(\tau_{\text{max}})}{I_{\text{rand}}(\tau_{\text{max}})}  
\end{equation}
with $I_{\tau_{\text{max}}}(\tau_{\text{max}})$ and $I_{\text{rand}}(\tau_{\text{max}})$ the intensity of the impulse response of the enhancing eigenstate and the time of flight at the delay time $\tau_{\text{max}}$. In Fig.~\ref{fig:enhancing_impulse_response}b., we compare the enhancement ratio $\eta(\tau)$ calculated with the experimentally measured impulse responses (red markers) with the expected enhancement $\eta^{\text{numerical}}(\tau)$ (green markers) for a set of different arrival times spread between $\tau=0$~ps and $\tau=45$~ps. The expected enhancement $\eta^{\text{numerical}}(\tau)$ is calculated with Eq.~\ref{eq:enhancement_ratio}, where $I_{\tau_{\text{max}}}^{\text{numerical}}$ replaces $I_{\tau_{\text{max}}}$. As the TRTM contains the full spatiotemporal relationship between the input and the output fields, $\eta(\tau)$ follows, as expected, a similar trend as $\eta^{\text{numerical}}(\tau)$.

\section{Attenuating the intensity over all spatial and polarization modes at an arbitrary delay time}

The time reversal operator $U_{\tau_s}^{\dagger} U_{\tau_s}$ enables to adjust the transmitted intensity over all spatial and polarization modes at the delay $\tau_s$. Exploiting the eigenvector associated with its maximum eigenvalue enables enhancement of intensity at $\tau_s$. Instead of enhancing the light delivery, we can also attenuate the impulse response at the delay time $\tau_s$ with the eigenvector $E_{\text{input}}^{\text{min}}$ associated with the minimum eigenvalue of $U_{\tau_s}^{\dagger} U_{\tau_s}$. We illustrate in Fig.~\ref{fig:extinction_impulse_response}a the experimental impulse response associated with the minimum eigenstate at a chosen delay time $\tau_{\text{min}}=19$~ps. The expected impulse response shows a clear minimum intensity at $\tau_{\text{min}}$. Experimentally, the impulse response reveals as well a sharp minimum at $\tau_{\text{min}}$, whose intensity is much smaller than the time of flight distribution at the same arrival time. 

Similarly to the enhancement ratio defined in Eq.~\ref{eq:enhancement_ratio}, we also define an attenuation ratio $\rho(\tau_{\text{min}})$ for such minimum eigenstates:
\begin{equation} \label{eq:extinction_ratio}
\rho(\tau_{\text{min}}) = \frac{I_{\text{rand}}(\tau_{\text{min}})}{I_{\tau_{\text{max}}}(\tau_{\text{min}})}  
\end{equation}
We plot the attenuation ratio in logarithm scale for a set of different delay time between 0~ps and 45~ps in Fig.~\ref{fig:extinction_impulse_response}b. Both the experimental and the expected attenuation ratios are following the same trend. However, the experimental conditions are not allowing a pure zero-intensity in the experimentally-measured impulse response. Indeed, the MSTM of the MMF has slightly decorrelated between its measurement and when the eigenstates are being measured. Besides, the superposition of the $N_{\text{input}}=254$ input modes on the SLM might not be as accurate as expected in the experiment. Finally, the experimental stability level, whether on the detection as well as on the illumination, are adding some noise which prevents the observation of a pure zero-intensity in the impulse response. 

\section{Polarization control of the generated output state}

The propagating modes in the multimode fiber are polarization-resolved. In order to address all the modes, it is essential to control both the spatial and the polarization degrees of freedom. Wavefront shaping experiments in disordered materials, including multimode fiber, are often measuring only a quarter of the transmission matrix, by neglecting a polarization state both at the input and at the output~\cite{popoff_measuring_2010,xiong_spatiotemporal_2016}. The experimental setup presented in Fig.~\ref{fig:exp_setup} enables the measurement of the polarization-resolved MSTM of the MMF. Indeed, we can generate and detect independently the two polarization states for all the spectral components. Therefore the MSTM $U(\lambda)$ could be written as follows:
\begin{equation}
U(\lambda)=
\begin{bmatrix}
    U_{HH}(\lambda)       & U_{HV}(\lambda) \\
   U_{VH}(\lambda)       & U_{VV}(\lambda) \\
    \end{bmatrix}
\end{equation}
with $U_{ij}(\lambda)$ the transmission matrix relating the input spatio-temporal field on the $i-th$ polarization state to the output spatio-temporal field on the $j-th$ polarization state. The dimension of each sub-matrix is $N_{\text{input}}/2 \times N_{\text{output}}/2$. Let's assume we want to generate an arbitrary state in a chosen polarization state $j$ at the output of the fiber at a wavelength $\lambda$, whether $j$ is H or V. For this purpose we thus use $U_j(\lambda)=[U_{Hj}(\lambda)~U_{Vj}(\lambda)]$~\cite{carpenter_complete_2016}. In the delay domain, the calculated TRTM from the experimentally measured MSTM is also polarization-resolved: $U_{j,\tau_s}$ is the TRTM at the delay time $\tau_s$ for the output polarization state $j$. The time reversal operator $U_{j,\tau_s}^{\dagger} U_{j,\tau_s}$ enables then a temporal control of the overall combination of spatial modes for a specific arrival time $\tau_s$ and a specific polarization state at the output of the MMF. 

\begin{figure}[htbp]
\centering
\includegraphics[width=\linewidth]{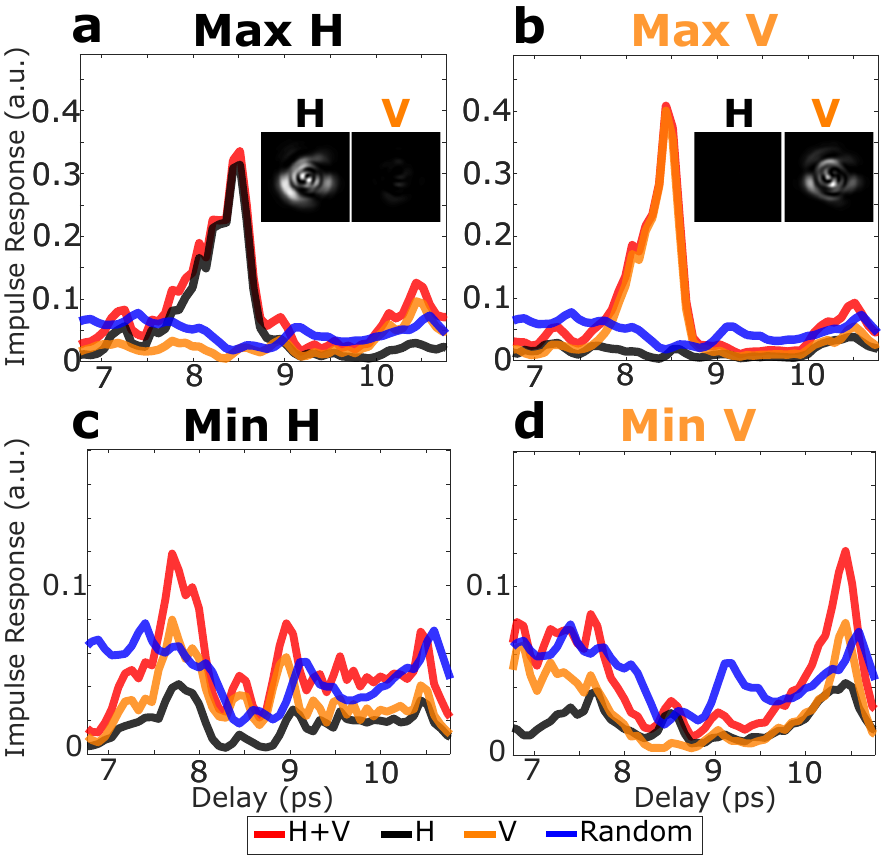}
\caption{Polarization control of the engineered state at an arbitrary delay time $\tau_s$. We choose here $\tau_s=8.5$~ps. (a) Enhancing only the H component of the output field. (b) Enhancing only the V component of the output field. (c) Attenuating only the H component of the output field. (d) Attenuating only the V component of the output field. Inset: reconstructed intensity at the output at delay time $\tau_s$ for both H and V. Color code: (red) Total impulse response of H and V; (black) Impulse response of the H component; (orange) Impulse response of the V component; (blue) Time of flight distribution of H and V.}
\label{fig:pola_control}
\end{figure}

Similarly to Fig.~\ref{fig:enhancing_impulse_response}, the output intensity can be enhanced over all the spatial modes for only the $j$-polarization state at the output. The input field $E_{\text{input}}^{\text{max,j}}$ is the eigenvector associated to the maximum eigenvalue of $U_{j,\tau_s}^{\dagger} U_{j,\tau_s}$. In Fig.~\ref{fig:pola_control}a and Fig.~\ref{fig:pola_control}b, we plot the polarization-resolved impulse responses associated to such eigenstates with $\tau_s=8.5$~ps. The polarization-resolved impulse responses are calculated with Eq.~\ref{eq:impulse_response}, using only the output modes on the chosen polarization state. Fig.~\ref{fig:pola_control}a and Fig.~\ref{fig:pola_control}b clearly evidence an enhancement of the intensity overall spatial modes at $\tau_s$ for either only H or only V. The reconstructed intensity at the output for H and V at the delay time $\tau_s$ confirms that the transmitted light at $\tau_s$ is polarized in the chosen state. The MMF in conjunction with the SLM can then be used as an accurate deterministic polarization control device in the time domain. This experiment could also be used to study polarization recovery in scattering media~\cite{aguiar_polarization_2017}.

We also demonstrate an attenuation of a chosen polarization state at the output of the fiber at the delay time $\tau_s$. Instead of launching $E_{\text{input}}^{\text{max,j}}$, we calculate $E_{\text{input}}^{\text{min,j}}$ which is the input eigenvector associated to the minimum eigenvalue of $U_{j,\tau_s}^{\dagger} U_{j,\tau_s}$. Fig.~\ref{fig:pola_control}c and Fig.~\ref{fig:pola_control}d present the experimental polarization-resolved impulse responses associated to such eigenstates for attenuating either only H or only V. Fig.~\ref{fig:pola_control}c and Fig.~\ref{fig:pola_control}d unambiguously reveal a strong attenuation at $\tau_s$ for the controlled output polarization state.

\section{Multiple delay control}

In Fig.~\ref{fig:enhancing_impulse_response}, we have demonstrated the enhancement of light intensity at a chosen arbitrary time $\tau_s$ over all the spatial and polarization modes. The TRTM could also be used to shape more complex impulse responses. Spatio-temporal focusing over multiple space-time positions have been demonstrated with digital optical phase conjugation of the TRTM~\cite{mounaix_deterministic_2016}. Here we demonstrate an enhancement of the transmitted light intensity at multiple delay times with the experimental setup of Fig.~\ref{fig:exp_setup}.

The time reversal operator $U_{\tau_s}^{\dagger} U_{\tau_s}$ enables to adjust the transmitted energy at the delay time $\tau_s$. Fig.~\ref{fig:enhancing_impulse_response} and Fig.~\ref{fig:extinction_impulse_response} have shown this control for either enhancing or attenuating light intensity at a chosen delay time $\tau_s$. The spectral diversity of the multimode fiber ensures that this control is only affecting delays around $\tau_s$, as the transmission matrices as a function of wavelength/delay are uncorrelated over a sufficient bandwidth/time interval~\cite{mounaix_spatiotemporal_2016,xiong_spatiotemporal_2016,carpenter_observation_2015}. 

In order to enhance light intensity at two different delays $\tau_1$ and $\tau_2$, we use the time reversal operators $U_{\tau_1}^{\dagger} U_{\tau_1}$ and $U_{\tau_2}^{\dagger} U_{\tau_2}$. We calculate the input eigenvectors $E_{\text{input}}^{\text{max},\tau_1}$ and $E_{\text{input}}^{\text{max},\tau_2}$ associated to the highest eigenvalue of  the two single operators. The linear superposition $E_{\text{input}}^{\text{max},\tau_1,\tau_2}= E_{\text{input}}^{\text{max},\tau_1} + E_{\text{input}}^{\text{max},\tau_2}$ enables a control of the impulse response at the two delay time $\tau_1$ and $\tau_2$. The phase of the solution would then be displayed on the SLM, similarly to spatio-temporal focusing at multiple delays~\cite{mounaix_deterministic_2016,mounaix_spatiotemporal_2016}. This control can be extended to a superposition of $N_{\text{superpos.}}$ delay times. The input field would then read:
\begin{equation} \label{eq:superposition_multiple}
E_{\text{input}}^{\text{max},N_{\text{superpos.}}} = \sum_{i=1}^{N_{\text{superpos.}}}  E_{\text{input}}^{\text{max},\tau_i}
\end{equation}
with $E_{\text{input}}^{\text{max},\tau_i}$ the eigenvector associated with the highest eigenvalue of $U_{\tau_i}^{\dagger} U_{\tau_i}$. In practice, $N_{\text{superpos.}}$ would be limited by the temporal width of the impulse response of each single mode. The impulse responses may interfere for different times, especially if the $\tau_i$ are close in the time domain. We plot in the Supplemental material the temporal correlation between the different Time Resolved Transmission Matrices. The quality of the superposition of the modes on the SLM would also limit the efficiency of a high $N_{\text{superpos.}}$.

\begin{figure}[htbp]
\centering
\includegraphics[width=\linewidth]{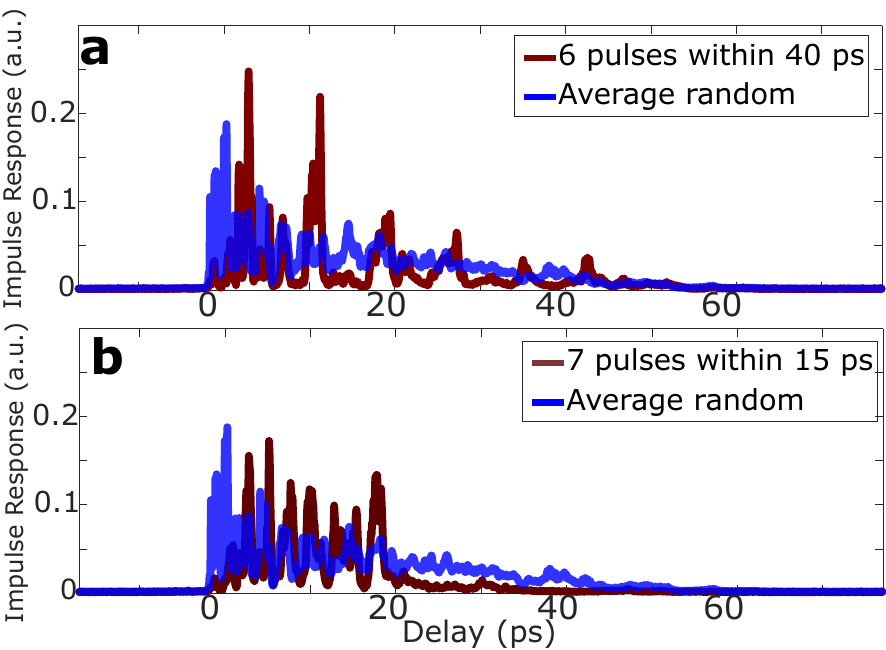}
\caption{Enhancement of transmitted light intensity at multiple delay times. (a) Enhancing the light intensity at 6 different delays {$\tau_i$} evenly spread from 4~ps to 44~ps. (b) Enhancing the light intensity at 7 different delays {$\tau_i$} evenly spread from 4~ps to 19~ps.}
\label{fig:multiple_delays}
\end{figure}

In Fig.~\ref{fig:multiple_delays} we demonstrate multiple delay control of the  transmitted intensity with the experimental setup of Fig.~\ref{fig:exp_setup}. In Fig.~\ref{fig:multiple_delays}a, $N_{\text{superpos.}}=6$ delay times are controlled, evenly spread between 4~ps and 44~ps.  Fig.~\ref{fig:multiple_delays}b illustrates a similar control over a narrower time interval. The intensity of $N_{\text{superpos.}}=7$ delay times is enhanced, evenly spread between 4~ps and 19~ps. We clearly identify $N_{\text{superpos.}}$ peaks in Fig.~\ref{fig:multiple_delays}a and Fig.~\ref{fig:multiple_delays}b. The amplitude ratio between these peaks could also be adjusted using an amplitude coefficient in the superposition in Eq.~\ref{eq:superposition_multiple}. A control of different $N_{\text{superpos.}}$ within the same time interval is presented in the Supplementary material. This multiple delay control could also be exploited in conjunction with Fig.~\ref{fig:pola_control}, where the two polarization states could be manipulated at different arrival times.

The ability to manipulate multiple delay times within long delay and dispersion provided by a very compact optical system such as a multimode fiber opens interesting perspectives for compact pulse shaping experiments. We believe this multiple delay control might also be useful for coherent control and pump probe experiments for imaging purposes or for complex control of light-matter interactions.

\section{Conclusion and perspectives}

Over the last decade, wavefront shaping with spatial light modulators have revolutionized the study of light propagation in disordered systems, where scattering was considered as an insurmountable obstacle. Most of the previous works  studied spatial control of the output field at the expense of the temporal behavior. We have demonstrated here the temporal control of light intensity averaged over all the spatial and polarization modes after propagation through a multimode fiber, at the expense of the spatial pattern. More precisely, we have shown how to enhance or to attenuate the transmitted intensity at any arbitrary delay times, arbitrary polarisation state, and arbitrary combinations of these. This work could be useful for the study of correlations in disordered systems~\cite{xiong_spatio-temporal_2018,hsu_correlation-enhanced_2017,freund_looking_1990,feng_correlations_1988} telecommunications~\cite{winzer_fiber-optic_2018}, non-linear imaging~\cite{de_aguiar_enhanced_2016,mounaix_transmission_2018} and endoscopy~\cite{papadopoulos_high-resolution_2013,caravaca-aguirre_single_2017, turtaev_high-fidelity_2018} through complex systems, time-gating imaging~\cite{badon_smart_2016,jeong_focusing_2018} and high power fiber laser~\cite{jackson_towards_2012}.

\paragraph{Funding Information}
Australian Research Council (ARC) projects DE180100009 and DP170101400.

\paragraph{Acknowledgements}
The authors thank Martin Pl\"oschner for careful reading of our manuscript.

\bibliographystyle{apsrev4-1}
\bibliography{TImeResolvedTMBiblio}

\begin{thebibliography}{53}%
\makeatletter
\providecommand \@ifxundefined [1]{%
 \@ifx{#1\undefined}
}%
\providecommand \@ifnum [1]{%
 \ifnum #1\expandafter \@firstoftwo
 \else \expandafter \@secondoftwo
 \fi
}%
\providecommand \@ifx [1]{%
 \ifx #1\expandafter \@firstoftwo
 \else \expandafter \@secondoftwo
 \fi
}%
\providecommand \natexlab [1]{#1}%
\providecommand \enquote  [1]{``#1''}%
\providecommand \bibnamefont  [1]{#1}%
\providecommand \bibfnamefont [1]{#1}%
\providecommand \citenamefont [1]{#1}%
\providecommand \href@noop [0]{\@secondoftwo}%
\providecommand \href [0]{\begingroup \@sanitize@url \@href}%
\providecommand \@href[1]{\@@startlink{#1}\@@href}%
\providecommand \@@href[1]{\endgroup#1\@@endlink}%
\providecommand \@sanitize@url [0]{\catcode `\\12\catcode `\$12\catcode
  `\&12\catcode `\#12\catcode `\^12\catcode `\_12\catcode `\%12\relax}%
\providecommand \@@startlink[1]{}%
\providecommand \@@endlink[0]{}%
\providecommand \url  [0]{\begingroup\@sanitize@url \@url }%
\providecommand \@url [1]{\endgroup\@href {#1}{\urlprefix }}%
\providecommand \urlprefix  [0]{URL }%
\providecommand \Eprint [0]{\href }%
\providecommand \doibase [0]{http://dx.doi.org/}%
\providecommand \selectlanguage [0]{\@gobble}%
\providecommand \bibinfo  [0]{\@secondoftwo}%
\providecommand \bibfield  [0]{\@secondoftwo}%
\providecommand \translation [1]{[#1]}%
\providecommand \BibitemOpen [0]{}%
\providecommand \bibitemStop [0]{}%
\providecommand \bibitemNoStop [0]{.\EOS\space}%
\providecommand \EOS [0]{\spacefactor3000\relax}%
\providecommand \BibitemShut  [1]{\csname bibitem#1\endcsname}%
\let\auto@bib@innerbib\@empty
\bibitem [{\citenamefont {Goodman}(1976)}]{goodman_fundamental_1976}%
  \BibitemOpen
  \bibfield  {author} {\bibinfo {author} {\bibfnamefont {J.~W.}\ \bibnamefont
  {Goodman}},\ }\href {\doibase 10.1364/JOSA.66.001145} {\bibfield  {journal}
  {\bibinfo  {journal} {JOSA}\ }\textbf {\bibinfo {volume} {66}},\ \bibinfo
  {pages} {1145} (\bibinfo {year} {1976})}\BibitemShut {NoStop}%
\bibitem [{\citenamefont {Rotter}\ and\ \citenamefont
  {Gigan}(2017)}]{rotter_light_2017}%
  \BibitemOpen
  \bibfield  {author} {\bibinfo {author} {\bibfnamefont {S.}~\bibnamefont
  {Rotter}}\ and\ \bibinfo {author} {\bibfnamefont {S.}~\bibnamefont {Gigan}},\
  }\href {\doibase 10.1103/RevModPhys.89.015005} {\bibfield  {journal}
  {\bibinfo  {journal} {Reviews of Modern Physics}\ }\textbf {\bibinfo {volume}
  {89}},\ \bibinfo {pages} {015005} (\bibinfo {year} {2017})}\BibitemShut
  {NoStop}%
\bibitem [{\citenamefont {Mosk}\ \emph {et~al.}(2012)\citenamefont {Mosk},
  \citenamefont {Lagendijk}, \citenamefont {Lerosey},\ and\ \citenamefont
  {Fink}}]{mosk_controlling_2012}%
  \BibitemOpen
  \bibfield  {author} {\bibinfo {author} {\bibfnamefont {A.~P.}\ \bibnamefont
  {Mosk}}, \bibinfo {author} {\bibfnamefont {A.}~\bibnamefont {Lagendijk}},
  \bibinfo {author} {\bibfnamefont {G.}~\bibnamefont {Lerosey}}, \ and\
  \bibinfo {author} {\bibfnamefont {M.}~\bibnamefont {Fink}},\ }\href {\doibase
  10.1038/nphoton.2012.88} {\bibfield  {journal} {\bibinfo  {journal} {Nature
  Photonics}\ }\textbf {\bibinfo {volume} {6}},\ \bibinfo {pages} {283}
  (\bibinfo {year} {2012})}\BibitemShut {NoStop}%
\bibitem [{\citenamefont {Vellekoop}\ and\ \citenamefont
  {Mosk}(2007)}]{vellekoop_focusing_2007}%
  \BibitemOpen
  \bibfield  {author} {\bibinfo {author} {\bibfnamefont {I.~M.}\ \bibnamefont
  {Vellekoop}}\ and\ \bibinfo {author} {\bibfnamefont {A.~P.}\ \bibnamefont
  {Mosk}},\ }\href {\doibase 10.1364/OL.32.002309} {\bibfield  {journal}
  {\bibinfo  {journal} {Optics Letters}\ }\textbf {\bibinfo {volume} {32}},\
  \bibinfo {pages} {2309} (\bibinfo {year} {2007})}\BibitemShut {NoStop}%
\bibitem [{\citenamefont {Vellekoop}(2015)}]{vellekoop_feedback-based_2015}%
  \BibitemOpen
  \bibfield  {author} {\bibinfo {author} {\bibfnamefont {I.~M.}\ \bibnamefont
  {Vellekoop}},\ }\href {\doibase 10.1364/OE.23.012189} {\bibfield  {journal}
  {\bibinfo  {journal} {Optics Express}\ }\textbf {\bibinfo {volume} {23}},\
  \bibinfo {pages} {12189} (\bibinfo {year} {2015})}\BibitemShut {NoStop}%
\bibitem [{\citenamefont {Yaqoob}\ \emph {et~al.}(2008)\citenamefont {Yaqoob},
  \citenamefont {Psaltis}, \citenamefont {Feld},\ and\ \citenamefont
  {Yang}}]{yaqoob_optical_2008}%
  \BibitemOpen
  \bibfield  {author} {\bibinfo {author} {\bibfnamefont {Z.}~\bibnamefont
  {Yaqoob}}, \bibinfo {author} {\bibfnamefont {D.}~\bibnamefont {Psaltis}},
  \bibinfo {author} {\bibfnamefont {M.~S.}\ \bibnamefont {Feld}}, \ and\
  \bibinfo {author} {\bibfnamefont {C.}~\bibnamefont {Yang}},\ }\href {\doibase
  10.1038/nphoton.2007.297} {\bibfield  {journal} {\bibinfo  {journal} {Nature
  Photonics}\ }\textbf {\bibinfo {volume} {2}},\ \bibinfo {pages} {110}
  (\bibinfo {year} {2008})}\BibitemShut {NoStop}%
\bibitem [{\citenamefont {Popoff}\ \emph
  {et~al.}(2010{\natexlab{a}})\citenamefont {Popoff}, \citenamefont {Lerosey},
  \citenamefont {Carminati}, \citenamefont {Fink}, \citenamefont {Boccara},\
  and\ \citenamefont {Gigan}}]{popoff_measuring_2010}%
  \BibitemOpen
  \bibfield  {author} {\bibinfo {author} {\bibfnamefont {S.~M.}\ \bibnamefont
  {Popoff}}, \bibinfo {author} {\bibfnamefont {G.}~\bibnamefont {Lerosey}},
  \bibinfo {author} {\bibfnamefont {R.}~\bibnamefont {Carminati}}, \bibinfo
  {author} {\bibfnamefont {M.}~\bibnamefont {Fink}}, \bibinfo {author}
  {\bibfnamefont {A.~C.}\ \bibnamefont {Boccara}}, \ and\ \bibinfo {author}
  {\bibfnamefont {S.}~\bibnamefont {Gigan}},\ }\href {\doibase
  10.1103/PhysRevLett.104.100601} {\bibfield  {journal} {\bibinfo  {journal}
  {Physical Review Letters}\ }\textbf {\bibinfo {volume} {104}},\ \bibinfo
  {pages} {100601} (\bibinfo {year} {2010}{\natexlab{a}})}\BibitemShut
  {NoStop}%
\bibitem [{\citenamefont {Guan}\ \emph {et~al.}(2012)\citenamefont {Guan},
  \citenamefont {Katz}, \citenamefont {Small}, \citenamefont {Zhou},\ and\
  \citenamefont {Silberberg}}]{guan_polarization_2012}%
  \BibitemOpen
  \bibfield  {author} {\bibinfo {author} {\bibfnamefont {Y.}~\bibnamefont
  {Guan}}, \bibinfo {author} {\bibfnamefont {O.}~\bibnamefont {Katz}}, \bibinfo
  {author} {\bibfnamefont {E.}~\bibnamefont {Small}}, \bibinfo {author}
  {\bibfnamefont {J.}~\bibnamefont {Zhou}}, \ and\ \bibinfo {author}
  {\bibfnamefont {Y.}~\bibnamefont {Silberberg}},\ }\href {\doibase
  10.1364/OL.37.004663} {\bibfield  {journal} {\bibinfo  {journal} {Optics
  Letters}\ }\textbf {\bibinfo {volume} {37}},\ \bibinfo {pages} {4663}
  (\bibinfo {year} {2012})}\BibitemShut {NoStop}%
\bibitem [{\citenamefont {Andreoli}\ \emph {et~al.}(2015)\citenamefont
  {Andreoli}, \citenamefont {Volpe}, \citenamefont {Popoff}, \citenamefont
  {Katz}, \citenamefont {Grésillon},\ and\ \citenamefont
  {Gigan}}]{andreoli_deterministic_2015}%
  \BibitemOpen
  \bibfield  {author} {\bibinfo {author} {\bibfnamefont {D.}~\bibnamefont
  {Andreoli}}, \bibinfo {author} {\bibfnamefont {G.}~\bibnamefont {Volpe}},
  \bibinfo {author} {\bibfnamefont {S.}~\bibnamefont {Popoff}}, \bibinfo
  {author} {\bibfnamefont {O.}~\bibnamefont {Katz}}, \bibinfo {author}
  {\bibfnamefont {S.}~\bibnamefont {Grésillon}}, \ and\ \bibinfo {author}
  {\bibfnamefont {S.}~\bibnamefont {Gigan}},\ }\href {\doibase
  10.1038/srep10347} {\bibfield  {journal} {\bibinfo  {journal} {Scientific
  Reports}\ }\textbf {\bibinfo {volume} {5}},\ \bibinfo {pages} {10347}
  (\bibinfo {year} {2015})}\BibitemShut {NoStop}%
\bibitem [{\citenamefont {Paudel}\ \emph {et~al.}(2013)\citenamefont {Paudel},
  \citenamefont {Stockbridge}, \citenamefont {Mertz},\ and\ \citenamefont
  {Bifano}}]{paudel_focusing_2013}%
  \BibitemOpen
  \bibfield  {author} {\bibinfo {author} {\bibfnamefont {H.~P.}\ \bibnamefont
  {Paudel}}, \bibinfo {author} {\bibfnamefont {C.}~\bibnamefont {Stockbridge}},
  \bibinfo {author} {\bibfnamefont {J.}~\bibnamefont {Mertz}}, \ and\ \bibinfo
  {author} {\bibfnamefont {T.}~\bibnamefont {Bifano}},\ }\href {\doibase
  10.1364/OE.21.017299} {\bibfield  {journal} {\bibinfo  {journal} {Optics
  Express}\ }\textbf {\bibinfo {volume} {21}},\ \bibinfo {pages} {17299}
  (\bibinfo {year} {2013})}\BibitemShut {NoStop}%
\bibitem [{\citenamefont {Mounaix}\ \emph
  {et~al.}(2016{\natexlab{a}})\citenamefont {Mounaix}, \citenamefont
  {Defienne},\ and\ \citenamefont {Gigan}}]{mounaix_deterministic_2016}%
  \BibitemOpen
  \bibfield  {author} {\bibinfo {author} {\bibfnamefont {M.}~\bibnamefont
  {Mounaix}}, \bibinfo {author} {\bibfnamefont {H.}~\bibnamefont {Defienne}}, \
  and\ \bibinfo {author} {\bibfnamefont {S.}~\bibnamefont {Gigan}},\ }\href
  {\doibase 10.1103/PhysRevA.94.041802} {\bibfield  {journal} {\bibinfo
  {journal} {Physical Review A}\ }\textbf {\bibinfo {volume} {94}},\ \bibinfo
  {pages} {041802} (\bibinfo {year} {2016}{\natexlab{a}})}\BibitemShut
  {NoStop}%
\bibitem [{\citenamefont {Čižmár}\ and\ \citenamefont
  {Dholakia}(2012)}]{cizmar_exploiting_2012}%
  \BibitemOpen
  \bibfield  {author} {\bibinfo {author} {\bibfnamefont {T.}~\bibnamefont
  {Čižmár}}\ and\ \bibinfo {author} {\bibfnamefont {K.}~\bibnamefont
  {Dholakia}},\ }\href {\doibase 10.1038/ncomms2024} {\bibfield  {journal}
  {\bibinfo  {journal} {Nature Communications}\ }\textbf {\bibinfo {volume}
  {3}},\ \bibinfo {pages} {1027} (\bibinfo {year} {2012})}\BibitemShut
  {NoStop}%
\bibitem [{\citenamefont {Choi}\ \emph {et~al.}(2012)\citenamefont {Choi},
  \citenamefont {Yoon}, \citenamefont {Kim}, \citenamefont {Yang},
  \citenamefont {Fang-Yen}, \citenamefont {Dasari}, \citenamefont {Lee},\ and\
  \citenamefont {Choi}}]{choi_scanner-free_2012}%
  \BibitemOpen
  \bibfield  {author} {\bibinfo {author} {\bibfnamefont {Y.}~\bibnamefont
  {Choi}}, \bibinfo {author} {\bibfnamefont {C.}~\bibnamefont {Yoon}}, \bibinfo
  {author} {\bibfnamefont {M.}~\bibnamefont {Kim}}, \bibinfo {author}
  {\bibfnamefont {T.~D.}\ \bibnamefont {Yang}}, \bibinfo {author}
  {\bibfnamefont {C.}~\bibnamefont {Fang-Yen}}, \bibinfo {author}
  {\bibfnamefont {R.~R.}\ \bibnamefont {Dasari}}, \bibinfo {author}
  {\bibfnamefont {K.~J.}\ \bibnamefont {Lee}}, \ and\ \bibinfo {author}
  {\bibfnamefont {W.}~\bibnamefont {Choi}},\ }\href {\doibase
  10.1103/PhysRevLett.109.203901} {\bibfield  {journal} {\bibinfo  {journal}
  {Physical Review Letters}\ }\textbf {\bibinfo {volume} {109}},\ \bibinfo
  {pages} {203901} (\bibinfo {year} {2012})}\BibitemShut {NoStop}%
\bibitem [{\citenamefont {Plöschner}\ \emph {et~al.}(2015)\citenamefont
  {Plöschner}, \citenamefont {Tyc},\ and\ \citenamefont
  {Čižmár}}]{ploschner_seeing_2015}%
  \BibitemOpen
  \bibfield  {author} {\bibinfo {author} {\bibfnamefont {M.}~\bibnamefont
  {Plöschner}}, \bibinfo {author} {\bibfnamefont {T.}~\bibnamefont {Tyc}}, \
  and\ \bibinfo {author} {\bibfnamefont {T.}~\bibnamefont {Čižmár}},\ }\href
  {\doibase 10.1038/nphoton.2015.112} {\bibfield  {journal} {\bibinfo
  {journal} {Nature Photonics}\ }\textbf {\bibinfo {volume} {9}},\ \bibinfo
  {pages} {529} (\bibinfo {year} {2015})}\BibitemShut {NoStop}%
\bibitem [{\citenamefont {Boonzajer~Flaes}\ \emph {et~al.}(2018)\citenamefont
  {Boonzajer~Flaes}, \citenamefont {Stopka}, \citenamefont {Turtaev},
  \citenamefont {de~Boer}, \citenamefont {Tyc},\ and\ \citenamefont
  {Čižmár}}]{boonzajer_flaes_robustness_2018}%
  \BibitemOpen
  \bibfield  {author} {\bibinfo {author} {\bibfnamefont {D.~E.}\ \bibnamefont
  {Boonzajer~Flaes}}, \bibinfo {author} {\bibfnamefont {J.}~\bibnamefont
  {Stopka}}, \bibinfo {author} {\bibfnamefont {S.}~\bibnamefont {Turtaev}},
  \bibinfo {author} {\bibfnamefont {J.~F.}\ \bibnamefont {de~Boer}}, \bibinfo
  {author} {\bibfnamefont {T.}~\bibnamefont {Tyc}}, \ and\ \bibinfo {author}
  {\bibfnamefont {T.}~\bibnamefont {Čižmár}},\ }\href {\doibase
  10.1103/PhysRevLett.120.233901} {\bibfield  {journal} {\bibinfo  {journal}
  {Physical Review Letters}\ }\textbf {\bibinfo {volume} {120}},\ \bibinfo
  {pages} {233901} (\bibinfo {year} {2018})}\BibitemShut {NoStop}%
\bibitem [{\citenamefont {Leite}\ \emph {et~al.}(2018)\citenamefont {Leite},
  \citenamefont {Turtaev}, \citenamefont {Jiang}, \citenamefont {Šiler},
  \citenamefont {Cuschieri}, \citenamefont {Russell},\ and\ \citenamefont
  {Čižmár}}]{leite_three-dimensional_2018}%
  \BibitemOpen
  \bibfield  {author} {\bibinfo {author} {\bibfnamefont {I.~T.}\ \bibnamefont
  {Leite}}, \bibinfo {author} {\bibfnamefont {S.}~\bibnamefont {Turtaev}},
  \bibinfo {author} {\bibfnamefont {X.}~\bibnamefont {Jiang}}, \bibinfo
  {author} {\bibfnamefont {M.}~\bibnamefont {Šiler}}, \bibinfo {author}
  {\bibfnamefont {A.}~\bibnamefont {Cuschieri}}, \bibinfo {author}
  {\bibfnamefont {P.~S.~J.}\ \bibnamefont {Russell}}, \ and\ \bibinfo {author}
  {\bibfnamefont {T.}~\bibnamefont {Čižmár}},\ }\href {\doibase
  10.1038/s41566-017-0053-8} {\bibfield  {journal} {\bibinfo  {journal} {Nature
  Photonics}\ }\textbf {\bibinfo {volume} {12}},\ \bibinfo {pages} {33}
  (\bibinfo {year} {2018})}\BibitemShut {NoStop}%
\bibitem [{\citenamefont {Aulbach}\ \emph {et~al.}(2011)\citenamefont
  {Aulbach}, \citenamefont {Gjonaj}, \citenamefont {Johnson}, \citenamefont
  {Mosk},\ and\ \citenamefont {Lagendijk}}]{aulbach_control_2011}%
  \BibitemOpen
  \bibfield  {author} {\bibinfo {author} {\bibfnamefont {J.}~\bibnamefont
  {Aulbach}}, \bibinfo {author} {\bibfnamefont {B.}~\bibnamefont {Gjonaj}},
  \bibinfo {author} {\bibfnamefont {P.~M.}\ \bibnamefont {Johnson}}, \bibinfo
  {author} {\bibfnamefont {A.~P.}\ \bibnamefont {Mosk}}, \ and\ \bibinfo
  {author} {\bibfnamefont {A.}~\bibnamefont {Lagendijk}},\ }\href {\doibase
  10.1103/PhysRevLett.106.103901} {\bibfield  {journal} {\bibinfo  {journal}
  {Physical Review Letters}\ }\textbf {\bibinfo {volume} {106}},\ \bibinfo
  {pages} {103901} (\bibinfo {year} {2011})}\BibitemShut {NoStop}%
\bibitem [{\citenamefont {Katz}\ \emph {et~al.}(2011)\citenamefont {Katz},
  \citenamefont {Small}, \citenamefont {Bromberg},\ and\ \citenamefont
  {Silberberg}}]{katz_focusing_2011}%
  \BibitemOpen
  \bibfield  {author} {\bibinfo {author} {\bibfnamefont {O.}~\bibnamefont
  {Katz}}, \bibinfo {author} {\bibfnamefont {E.}~\bibnamefont {Small}},
  \bibinfo {author} {\bibfnamefont {Y.}~\bibnamefont {Bromberg}}, \ and\
  \bibinfo {author} {\bibfnamefont {Y.}~\bibnamefont {Silberberg}},\ }\href
  {\doibase 10.1038/nphoton.2011.72} {\bibfield  {journal} {\bibinfo  {journal}
  {Nature Photonics}\ }\textbf {\bibinfo {volume} {5}},\ \bibinfo {pages} {372}
  (\bibinfo {year} {2011})}\BibitemShut {NoStop}%
\bibitem [{\citenamefont {Morales-Delgado}\ \emph {et~al.}(2015)\citenamefont
  {Morales-Delgado}, \citenamefont {Farahi}, \citenamefont {Papadopoulos},
  \citenamefont {Psaltis},\ and\ \citenamefont
  {Moser}}]{morales-delgado_delivery_2015}%
  \BibitemOpen
  \bibfield  {author} {\bibinfo {author} {\bibfnamefont {E.~E.}\ \bibnamefont
  {Morales-Delgado}}, \bibinfo {author} {\bibfnamefont {S.}~\bibnamefont
  {Farahi}}, \bibinfo {author} {\bibfnamefont {I.~N.}\ \bibnamefont
  {Papadopoulos}}, \bibinfo {author} {\bibfnamefont {D.}~\bibnamefont
  {Psaltis}}, \ and\ \bibinfo {author} {\bibfnamefont {C.}~\bibnamefont
  {Moser}},\ }\href {\doibase 10.1364/OE.23.009109} {\bibfield  {journal}
  {\bibinfo  {journal} {Optics Express}\ }\textbf {\bibinfo {volume} {23}},\
  \bibinfo {pages} {9109} (\bibinfo {year} {2015})}\BibitemShut {NoStop}%
\bibitem [{\citenamefont {McCabe}\ \emph {et~al.}(2011)\citenamefont {McCabe},
  \citenamefont {Tajalli}, \citenamefont {Austin}, \citenamefont {Bondareff},
  \citenamefont {Walmsley}, \citenamefont {Gigan},\ and\ \citenamefont
  {Chatel}}]{mccabe_spatio-temporal_2011}%
  \BibitemOpen
  \bibfield  {author} {\bibinfo {author} {\bibfnamefont {D.~J.}\ \bibnamefont
  {McCabe}}, \bibinfo {author} {\bibfnamefont {A.}~\bibnamefont {Tajalli}},
  \bibinfo {author} {\bibfnamefont {D.~R.}\ \bibnamefont {Austin}}, \bibinfo
  {author} {\bibfnamefont {P.}~\bibnamefont {Bondareff}}, \bibinfo {author}
  {\bibfnamefont {I.~A.}\ \bibnamefont {Walmsley}}, \bibinfo {author}
  {\bibfnamefont {S.}~\bibnamefont {Gigan}}, \ and\ \bibinfo {author}
  {\bibfnamefont {B.}~\bibnamefont {Chatel}},\ }\href {\doibase
  10.1038/ncomms1434} {\bibfield  {journal} {\bibinfo  {journal} {Nature
  Communications}\ }\textbf {\bibinfo {volume} {2}},\ \bibinfo {pages} {447}
  (\bibinfo {year} {2011})}\BibitemShut {NoStop}%
\bibitem [{\citenamefont {Mounaix}\ \emph
  {et~al.}(2016{\natexlab{b}})\citenamefont {Mounaix}, \citenamefont
  {Andreoli}, \citenamefont {Defienne}, \citenamefont {Volpe}, \citenamefont
  {Katz}, \citenamefont {Grésillon},\ and\ \citenamefont
  {Gigan}}]{mounaix_spatiotemporal_2016}%
  \BibitemOpen
  \bibfield  {author} {\bibinfo {author} {\bibfnamefont {M.}~\bibnamefont
  {Mounaix}}, \bibinfo {author} {\bibfnamefont {D.}~\bibnamefont {Andreoli}},
  \bibinfo {author} {\bibfnamefont {H.}~\bibnamefont {Defienne}}, \bibinfo
  {author} {\bibfnamefont {G.}~\bibnamefont {Volpe}}, \bibinfo {author}
  {\bibfnamefont {O.}~\bibnamefont {Katz}}, \bibinfo {author} {\bibfnamefont
  {S.}~\bibnamefont {Grésillon}}, \ and\ \bibinfo {author} {\bibfnamefont
  {S.}~\bibnamefont {Gigan}},\ }\href {\doibase 10.1103/PhysRevLett.116.253901}
  {\bibfield  {journal} {\bibinfo  {journal} {Physical Review Letters}\
  }\textbf {\bibinfo {volume} {116}},\ \bibinfo {pages} {253901} (\bibinfo
  {year} {2016}{\natexlab{b}})}\BibitemShut {NoStop}%
\bibitem [{\citenamefont {Mounaix}\ \emph {et~al.}(2017)\citenamefont
  {Mounaix}, \citenamefont {Aguiar},\ and\ \citenamefont
  {Gigan}}]{mounaix_temporal_2017}%
  \BibitemOpen
  \bibfield  {author} {\bibinfo {author} {\bibfnamefont {M.}~\bibnamefont
  {Mounaix}}, \bibinfo {author} {\bibfnamefont {H.~B.~d.}\ \bibnamefont
  {Aguiar}}, \ and\ \bibinfo {author} {\bibfnamefont {S.}~\bibnamefont
  {Gigan}},\ }\href {\doibase 10.1364/OPTICA.4.001289} {\bibfield  {journal}
  {\bibinfo  {journal} {Optica}\ }\textbf {\bibinfo {volume} {4}},\ \bibinfo
  {pages} {1289} (\bibinfo {year} {2017})}\BibitemShut {NoStop}%
\bibitem [{\citenamefont {Carpenter}\ \emph {et~al.}(2015)\citenamefont
  {Carpenter}, \citenamefont {Eggleton},\ and\ \citenamefont
  {Schröder}}]{carpenter_observation_2015}%
  \BibitemOpen
  \bibfield  {author} {\bibinfo {author} {\bibfnamefont {J.}~\bibnamefont
  {Carpenter}}, \bibinfo {author} {\bibfnamefont {B.~J.}\ \bibnamefont
  {Eggleton}}, \ and\ \bibinfo {author} {\bibfnamefont {J.}~\bibnamefont
  {Schröder}},\ }\href {\doibase 10.1038/nphoton.2015.188} {\bibfield
  {journal} {\bibinfo  {journal} {Nature Photonics}\ }\textbf {\bibinfo
  {volume} {9}},\ \bibinfo {pages} {751} (\bibinfo {year} {2015})}\BibitemShut
  {NoStop}%
\bibitem [{\citenamefont {Xiong}\ \emph {et~al.}(2016)\citenamefont {Xiong},
  \citenamefont {Ambichl}, \citenamefont {Bromberg}, \citenamefont {Redding},
  \citenamefont {Rotter},\ and\ \citenamefont
  {Cao}}]{xiong_spatiotemporal_2016}%
  \BibitemOpen
  \bibfield  {author} {\bibinfo {author} {\bibfnamefont {W.}~\bibnamefont
  {Xiong}}, \bibinfo {author} {\bibfnamefont {P.}~\bibnamefont {Ambichl}},
  \bibinfo {author} {\bibfnamefont {Y.}~\bibnamefont {Bromberg}}, \bibinfo
  {author} {\bibfnamefont {B.}~\bibnamefont {Redding}}, \bibinfo {author}
  {\bibfnamefont {S.}~\bibnamefont {Rotter}}, \ and\ \bibinfo {author}
  {\bibfnamefont {H.}~\bibnamefont {Cao}},\ }\href {\doibase
  10.1103/PhysRevLett.117.053901} {\bibfield  {journal} {\bibinfo  {journal}
  {Physical Review Letters}\ }\textbf {\bibinfo {volume} {117}},\ \bibinfo
  {pages} {053901} (\bibinfo {year} {2016})}\BibitemShut {NoStop}%
\bibitem [{\citenamefont {Ambichl}\ \emph {et~al.}(2017)\citenamefont
  {Ambichl}, \citenamefont {Xiong}, \citenamefont {Bromberg}, \citenamefont
  {Redding}, \citenamefont {Cao},\ and\ \citenamefont
  {Rotter}}]{ambichl_super-_2017-1}%
  \BibitemOpen
  \bibfield  {author} {\bibinfo {author} {\bibfnamefont {P.}~\bibnamefont
  {Ambichl}}, \bibinfo {author} {\bibfnamefont {W.}~\bibnamefont {Xiong}},
  \bibinfo {author} {\bibfnamefont {Y.}~\bibnamefont {Bromberg}}, \bibinfo
  {author} {\bibfnamefont {B.}~\bibnamefont {Redding}}, \bibinfo {author}
  {\bibfnamefont {H.}~\bibnamefont {Cao}}, \ and\ \bibinfo {author}
  {\bibfnamefont {S.}~\bibnamefont {Rotter}},\ }\href {\doibase
  10.1103/PhysRevX.7.041053} {\bibfield  {journal} {\bibinfo  {journal}
  {Physical Review X}\ }\textbf {\bibinfo {volume} {7}},\ \bibinfo {pages}
  {041053} (\bibinfo {year} {2017})}\BibitemShut {NoStop}%
\bibitem [{\citenamefont {Popoff}\ \emph
  {et~al.}(2010{\natexlab{b}})\citenamefont {Popoff}, \citenamefont {Lerosey},
  \citenamefont {Fink}, \citenamefont {Boccara},\ and\ \citenamefont
  {Gigan}}]{popoff_image_2010}%
  \BibitemOpen
  \bibfield  {author} {\bibinfo {author} {\bibfnamefont {S.}~\bibnamefont
  {Popoff}}, \bibinfo {author} {\bibfnamefont {G.}~\bibnamefont {Lerosey}},
  \bibinfo {author} {\bibfnamefont {M.}~\bibnamefont {Fink}}, \bibinfo {author}
  {\bibfnamefont {A.~C.}\ \bibnamefont {Boccara}}, \ and\ \bibinfo {author}
  {\bibfnamefont {S.}~\bibnamefont {Gigan}},\ }\href {\doibase
  10.1038/ncomms1078} {\bibfield  {journal} {\bibinfo  {journal} {Nature
  Communications}\ }\textbf {\bibinfo {volume} {1}},\ \bibinfo {pages} {81}
  (\bibinfo {year} {2010}{\natexlab{b}})}\BibitemShut {NoStop}%
\bibitem [{\citenamefont {Carpenter}\ \emph {et~al.}(2016)\citenamefont
  {Carpenter}, \citenamefont {Eggleton},\ and\ \citenamefont
  {Schröder}}]{carpenter_complete_2016}%
  \BibitemOpen
  \bibfield  {author} {\bibinfo {author} {\bibfnamefont {J.}~\bibnamefont
  {Carpenter}}, \bibinfo {author} {\bibfnamefont {B.~J.}\ \bibnamefont
  {Eggleton}}, \ and\ \bibinfo {author} {\bibfnamefont {J.}~\bibnamefont
  {Schröder}},\ }\href {\doibase 10.1364/OL.41.005580} {\bibfield  {journal}
  {\bibinfo  {journal} {Optics Letters}\ }\textbf {\bibinfo {volume} {41}},\
  \bibinfo {pages} {5580} (\bibinfo {year} {2016})}\BibitemShut {NoStop}%
\bibitem [{\citenamefont {Boniface}\ \emph {et~al.}(2019)\citenamefont
  {Boniface}, \citenamefont {Gusachenko}, \citenamefont {Dholakia},\ and\
  \citenamefont {Gigan}}]{boniface_rapid_2019}%
  \BibitemOpen
  \bibfield  {author} {\bibinfo {author} {\bibfnamefont {A.}~\bibnamefont
  {Boniface}}, \bibinfo {author} {\bibfnamefont {I.}~\bibnamefont
  {Gusachenko}}, \bibinfo {author} {\bibfnamefont {K.}~\bibnamefont
  {Dholakia}}, \ and\ \bibinfo {author} {\bibfnamefont {S.}~\bibnamefont
  {Gigan}},\ }\href {\doibase 10.1364/OPTICA.6.000274} {\bibfield  {journal}
  {\bibinfo  {journal} {Optica}\ }\textbf {\bibinfo {volume} {6}},\ \bibinfo
  {pages} {274} (\bibinfo {year} {2019})}\BibitemShut {NoStop}%
\bibitem [{\citenamefont {Mounaix}\ \emph {et~al.}(2018)\citenamefont
  {Mounaix}, \citenamefont {Ta},\ and\ \citenamefont
  {Gigan}}]{mounaix_transmission_2018}%
  \BibitemOpen
  \bibfield  {author} {\bibinfo {author} {\bibfnamefont {M.}~\bibnamefont
  {Mounaix}}, \bibinfo {author} {\bibfnamefont {D.~M.}\ \bibnamefont {Ta}}, \
  and\ \bibinfo {author} {\bibfnamefont {S.}~\bibnamefont {Gigan}},\ }\href
  {\doibase 10.1364/OL.43.002831} {\bibfield  {journal} {\bibinfo  {journal}
  {Optics Letters}\ }\textbf {\bibinfo {volume} {43}},\ \bibinfo {pages} {2831}
  (\bibinfo {year} {2018})}\BibitemShut {NoStop}%
\bibitem [{\citenamefont {Carpenter}\ \emph {et~al.}(2014)\citenamefont
  {Carpenter}, \citenamefont {Eggleton},\ and\ \citenamefont
  {Schröder}}]{carpenter_110x110_2014}%
  \BibitemOpen
  \bibfield  {author} {\bibinfo {author} {\bibfnamefont {J.}~\bibnamefont
  {Carpenter}}, \bibinfo {author} {\bibfnamefont {B.~J.}\ \bibnamefont
  {Eggleton}}, \ and\ \bibinfo {author} {\bibfnamefont {J.}~\bibnamefont
  {Schröder}},\ }\href {\doibase 10.1364/OE.22.000096} {\bibfield  {journal}
  {\bibinfo  {journal} {Optics Express}\ }\textbf {\bibinfo {volume} {22}},\
  \bibinfo {pages} {96} (\bibinfo {year} {2014})}\BibitemShut {NoStop}%
\bibitem [{\citenamefont {Snyder}\ and\ \citenamefont
  {Love}(2012)}]{snyder_optical_2012}%
  \BibitemOpen
  \bibfield  {author} {\bibinfo {author} {\bibfnamefont {A.~W.}\ \bibnamefont
  {Snyder}}\ and\ \bibinfo {author} {\bibfnamefont {J.}~\bibnamefont {Love}},\
  }\href@noop {} {\emph {\bibinfo {title} {Optical waveguide theory}}}\
  (\bibinfo  {publisher} {Springer Science \& Business Media},\ \bibinfo {year}
  {2012})\BibitemShut {NoStop}%
\bibitem [{\citenamefont {Xiong}\ \emph {et~al.}(2018)\citenamefont {Xiong},
  \citenamefont {Hsu},\ and\ \citenamefont {Cao}}]{xiong_spatio-temporal_2018}%
  \BibitemOpen
  \bibfield  {author} {\bibinfo {author} {\bibfnamefont {W.}~\bibnamefont
  {Xiong}}, \bibinfo {author} {\bibfnamefont {C.~W.}\ \bibnamefont {Hsu}}, \
  and\ \bibinfo {author} {\bibfnamefont {H.}~\bibnamefont {Cao}},\ }\href
  {http://arxiv.org/abs/1811.02552} {\bibfield  {journal} {\bibinfo  {journal}
  {arXiv:1811.02552 [physics]}\ } (\bibinfo {year} {2018})},\ \bibinfo {note}
  {arXiv: 1811.02552}\BibitemShut {NoStop}%
\bibitem [{\citenamefont {Prada}\ and\ \citenamefont
  {Fink}(1994)}]{prada_eigenmodes_1994}%
  \BibitemOpen
  \bibfield  {author} {\bibinfo {author} {\bibfnamefont {C.}~\bibnamefont
  {Prada}}\ and\ \bibinfo {author} {\bibfnamefont {M.}~\bibnamefont {Fink}},\
  }\href {\doibase 10.1016/0165-2125(94)90039-6} {\bibfield  {journal}
  {\bibinfo  {journal} {Wave Motion}\ }\textbf {\bibinfo {volume} {20}},\
  \bibinfo {pages} {151} (\bibinfo {year} {1994})}\BibitemShut {NoStop}%
\bibitem [{\citenamefont {Prada}\ \emph {et~al.}(1996)\citenamefont {Prada},
  \citenamefont {Manneville}, \citenamefont {Spoliansky},\ and\ \citenamefont
  {Fink}}]{prada_decomposition_1996}%
  \BibitemOpen
  \bibfield  {author} {\bibinfo {author} {\bibfnamefont {C.}~\bibnamefont
  {Prada}}, \bibinfo {author} {\bibfnamefont {S.}~\bibnamefont {Manneville}},
  \bibinfo {author} {\bibfnamefont {D.}~\bibnamefont {Spoliansky}}, \ and\
  \bibinfo {author} {\bibfnamefont {M.}~\bibnamefont {Fink}},\ }\href {\doibase
  10.1121/1.415393} {\bibfield  {journal} {\bibinfo  {journal} {The Journal of
  the Acoustical Society of America}\ }\textbf {\bibinfo {volume} {99}},\
  \bibinfo {pages} {2067} (\bibinfo {year} {1996})}\BibitemShut {NoStop}%
\bibitem [{\citenamefont {Popoff}\ \emph {et~al.}(2011)\citenamefont {Popoff},
  \citenamefont {Aubry}, \citenamefont {Lerosey}, \citenamefont {Fink},
  \citenamefont {Boccara},\ and\ \citenamefont
  {Gigan}}]{popoff_exploiting_2011}%
  \BibitemOpen
  \bibfield  {author} {\bibinfo {author} {\bibfnamefont {S.~M.}\ \bibnamefont
  {Popoff}}, \bibinfo {author} {\bibfnamefont {A.}~\bibnamefont {Aubry}},
  \bibinfo {author} {\bibfnamefont {G.}~\bibnamefont {Lerosey}}, \bibinfo
  {author} {\bibfnamefont {M.}~\bibnamefont {Fink}}, \bibinfo {author}
  {\bibfnamefont {A.~C.}\ \bibnamefont {Boccara}}, \ and\ \bibinfo {author}
  {\bibfnamefont {S.}~\bibnamefont {Gigan}},\ }\href {\doibase
  10.1103/PhysRevLett.107.263901} {\bibfield  {journal} {\bibinfo  {journal}
  {Physical Review Letters}\ }\textbf {\bibinfo {volume} {107}},\ \bibinfo
  {pages} {263901} (\bibinfo {year} {2011})}\BibitemShut {NoStop}%
\bibitem [{\citenamefont {Choi}\ \emph {et~al.}(2011)\citenamefont {Choi},
  \citenamefont {Mosk}, \citenamefont {Park},\ and\ \citenamefont
  {Choi}}]{choi_transmission_2011}%
  \BibitemOpen
  \bibfield  {author} {\bibinfo {author} {\bibfnamefont {W.}~\bibnamefont
  {Choi}}, \bibinfo {author} {\bibfnamefont {A.~P.}\ \bibnamefont {Mosk}},
  \bibinfo {author} {\bibfnamefont {Q.-H.}\ \bibnamefont {Park}}, \ and\
  \bibinfo {author} {\bibfnamefont {W.}~\bibnamefont {Choi}},\ }\href {\doibase
  10.1103/PhysRevB.83.134207} {\bibfield  {journal} {\bibinfo  {journal}
  {Physical Review B}\ }\textbf {\bibinfo {volume} {83}},\ \bibinfo {pages}
  {134207} (\bibinfo {year} {2011})}\BibitemShut {NoStop}%
\bibitem [{\citenamefont {Davy}\ \emph
  {et~al.}(2015{\natexlab{a}})\citenamefont {Davy}, \citenamefont {Shi},
  \citenamefont {Wang}, \citenamefont {Cheng},\ and\ \citenamefont
  {Genack}}]{davy_transmission_2015}%
  \BibitemOpen
  \bibfield  {author} {\bibinfo {author} {\bibfnamefont {M.}~\bibnamefont
  {Davy}}, \bibinfo {author} {\bibfnamefont {Z.}~\bibnamefont {Shi}}, \bibinfo
  {author} {\bibfnamefont {J.}~\bibnamefont {Wang}}, \bibinfo {author}
  {\bibfnamefont {X.}~\bibnamefont {Cheng}}, \ and\ \bibinfo {author}
  {\bibfnamefont {A.~Z.}\ \bibnamefont {Genack}},\ }\href {\doibase
  10.1103/PhysRevLett.114.033901} {\bibfield  {journal} {\bibinfo  {journal}
  {Physical Review Letters}\ }\textbf {\bibinfo {volume} {114}},\ \bibinfo
  {pages} {033901} (\bibinfo {year} {2015}{\natexlab{a}})}\BibitemShut
  {NoStop}%
\bibitem [{\citenamefont {Davy}\ \emph
  {et~al.}(2015{\natexlab{b}})\citenamefont {Davy}, \citenamefont {Shi},
  \citenamefont {Park}, \citenamefont {Tian},\ and\ \citenamefont
  {Genack}}]{davy_universal_2015}%
  \BibitemOpen
  \bibfield  {author} {\bibinfo {author} {\bibfnamefont {M.}~\bibnamefont
  {Davy}}, \bibinfo {author} {\bibfnamefont {Z.}~\bibnamefont {Shi}}, \bibinfo
  {author} {\bibfnamefont {J.}~\bibnamefont {Park}}, \bibinfo {author}
  {\bibfnamefont {C.}~\bibnamefont {Tian}}, \ and\ \bibinfo {author}
  {\bibfnamefont {A.~Z.}\ \bibnamefont {Genack}},\ }\href {\doibase
  10.1038/ncomms7893} {\bibfield  {journal} {\bibinfo  {journal} {Nature
  Communications}\ }\textbf {\bibinfo {volume} {6}},\ \bibinfo {pages} {6893}
  (\bibinfo {year} {2015}{\natexlab{b}})}\BibitemShut {NoStop}%
\bibitem [{\citenamefont {Yılmaz}\ \emph {et~al.}(2019)\citenamefont
  {Yılmaz}, \citenamefont {Hsu}, \citenamefont {Yamilov},\ and\ \citenamefont
  {Cao}}]{yilmaz_transverse_2019}%
  \BibitemOpen
  \bibfield  {author} {\bibinfo {author} {\bibfnamefont {H.}~\bibnamefont
  {Yılmaz}}, \bibinfo {author} {\bibfnamefont {C.~W.}\ \bibnamefont {Hsu}},
  \bibinfo {author} {\bibfnamefont {A.}~\bibnamefont {Yamilov}}, \ and\
  \bibinfo {author} {\bibfnamefont {H.}~\bibnamefont {Cao}},\ }\href {\doibase
  10.1038/s41566-019-0367-9} {\bibfield  {journal} {\bibinfo  {journal} {Nature
  Photonics}\ ,\ \bibinfo {pages} {1}} (\bibinfo {year} {2019})}\BibitemShut
  {NoStop}%
\bibitem [{\citenamefont {Kim}\ \emph {et~al.}(2012)\citenamefont {Kim},
  \citenamefont {Choi}, \citenamefont {Yoon}, \citenamefont {Choi},
  \citenamefont {Kim}, \citenamefont {Park},\ and\ \citenamefont
  {Choi}}]{kim_maximal_2012}%
  \BibitemOpen
  \bibfield  {author} {\bibinfo {author} {\bibfnamefont {M.}~\bibnamefont
  {Kim}}, \bibinfo {author} {\bibfnamefont {Y.}~\bibnamefont {Choi}}, \bibinfo
  {author} {\bibfnamefont {C.}~\bibnamefont {Yoon}}, \bibinfo {author}
  {\bibfnamefont {W.}~\bibnamefont {Choi}}, \bibinfo {author} {\bibfnamefont
  {J.}~\bibnamefont {Kim}}, \bibinfo {author} {\bibfnamefont {Q.-H.}\
  \bibnamefont {Park}}, \ and\ \bibinfo {author} {\bibfnamefont
  {W.}~\bibnamefont {Choi}},\ }\href {\doibase 10.1038/nphoton.2012.159}
  {\bibfield  {journal} {\bibinfo  {journal} {Nature Photonics}\ }\textbf
  {\bibinfo {volume} {6}},\ \bibinfo {pages} {581} (\bibinfo {year}
  {2012})}\BibitemShut {NoStop}%
\bibitem [{\citenamefont {Hsu}\ \emph {et~al.}(2015)\citenamefont {Hsu},
  \citenamefont {Goetschy}, \citenamefont {Bromberg}, \citenamefont {Stone},\
  and\ \citenamefont {Cao}}]{hsu_broadband_2015}%
  \BibitemOpen
  \bibfield  {author} {\bibinfo {author} {\bibfnamefont {C.~W.}\ \bibnamefont
  {Hsu}}, \bibinfo {author} {\bibfnamefont {A.}~\bibnamefont {Goetschy}},
  \bibinfo {author} {\bibfnamefont {Y.}~\bibnamefont {Bromberg}}, \bibinfo
  {author} {\bibfnamefont {A.~D.}\ \bibnamefont {Stone}}, \ and\ \bibinfo
  {author} {\bibfnamefont {H.}~\bibnamefont {Cao}},\ }\href {\doibase
  10.1103/PhysRevLett.115.223901} {\bibfield  {journal} {\bibinfo  {journal}
  {Physical Review Letters}\ }\textbf {\bibinfo {volume} {115}},\ \bibinfo
  {pages} {223901} (\bibinfo {year} {2015})}\BibitemShut {NoStop}%
\bibitem [{\citenamefont {Aguiar}\ \emph {et~al.}(2017)\citenamefont {Aguiar},
  \citenamefont {Gigan},\ and\ \citenamefont
  {Brasselet}}]{aguiar_polarization_2017}%
  \BibitemOpen
  \bibfield  {author} {\bibinfo {author} {\bibfnamefont {H.~B.~d.}\
  \bibnamefont {Aguiar}}, \bibinfo {author} {\bibfnamefont {S.}~\bibnamefont
  {Gigan}}, \ and\ \bibinfo {author} {\bibfnamefont {S.}~\bibnamefont
  {Brasselet}},\ }\href {\doibase 10.1126/sciadv.1600743} {\bibfield  {journal}
  {\bibinfo  {journal} {Science Advances}\ }\textbf {\bibinfo {volume} {3}},\
  \bibinfo {pages} {e1600743} (\bibinfo {year} {2017})}\BibitemShut {NoStop}%
\bibitem [{\citenamefont {Hsu}\ \emph {et~al.}(2017)\citenamefont {Hsu},
  \citenamefont {Liew}, \citenamefont {Goetschy}, \citenamefont {Cao},\ and\
  \citenamefont {Douglas~Stone}}]{hsu_correlation-enhanced_2017}%
  \BibitemOpen
  \bibfield  {author} {\bibinfo {author} {\bibfnamefont {C.~W.}\ \bibnamefont
  {Hsu}}, \bibinfo {author} {\bibfnamefont {S.~F.}\ \bibnamefont {Liew}},
  \bibinfo {author} {\bibfnamefont {A.}~\bibnamefont {Goetschy}}, \bibinfo
  {author} {\bibfnamefont {H.}~\bibnamefont {Cao}}, \ and\ \bibinfo {author}
  {\bibfnamefont {A.}~\bibnamefont {Douglas~Stone}},\ }\href {\doibase
  10.1038/nphys4036} {\bibfield  {journal} {\bibinfo  {journal} {Nature
  Physics}\ }\textbf {\bibinfo {volume} {advance online publication}} (\bibinfo
  {year} {2017}),\ 10.1038/nphys4036}\BibitemShut {NoStop}%
\bibitem [{\citenamefont {Freund}(1990)}]{freund_looking_1990}%
  \BibitemOpen
  \bibfield  {author} {\bibinfo {author} {\bibfnamefont {I.}~\bibnamefont
  {Freund}},\ }\href {\doibase 10.1016/0378-4371(90)90357-X} {\bibfield
  {journal} {\bibinfo  {journal} {Physica A: Statistical Mechanics and its
  Applications}\ }\textbf {\bibinfo {volume} {168}},\ \bibinfo {pages} {49}
  (\bibinfo {year} {1990})}\BibitemShut {NoStop}%
\bibitem [{\citenamefont {Feng}\ \emph {et~al.}(1988)\citenamefont {Feng},
  \citenamefont {Kane}, \citenamefont {Lee},\ and\ \citenamefont
  {Stone}}]{feng_correlations_1988}%
  \BibitemOpen
  \bibfield  {author} {\bibinfo {author} {\bibfnamefont {S.}~\bibnamefont
  {Feng}}, \bibinfo {author} {\bibfnamefont {C.}~\bibnamefont {Kane}}, \bibinfo
  {author} {\bibfnamefont {P.~A.}\ \bibnamefont {Lee}}, \ and\ \bibinfo
  {author} {\bibfnamefont {A.~D.}\ \bibnamefont {Stone}},\ }\href {\doibase
  10.1103/PhysRevLett.61.834} {\bibfield  {journal} {\bibinfo  {journal}
  {Physical Review Letters}\ }\textbf {\bibinfo {volume} {61}},\ \bibinfo
  {pages} {834} (\bibinfo {year} {1988})}\BibitemShut {NoStop}%
\bibitem [{\citenamefont {Winzer}\ \emph {et~al.}(2018)\citenamefont {Winzer},
  \citenamefont {Neilson},\ and\ \citenamefont
  {Chraplyvy}}]{winzer_fiber-optic_2018}%
  \BibitemOpen
  \bibfield  {author} {\bibinfo {author} {\bibfnamefont {P.~J.}\ \bibnamefont
  {Winzer}}, \bibinfo {author} {\bibfnamefont {D.~T.}\ \bibnamefont {Neilson}},
  \ and\ \bibinfo {author} {\bibfnamefont {A.~R.}\ \bibnamefont {Chraplyvy}},\
  }\href {\doibase 10.1364/OE.26.024190} {\bibfield  {journal} {\bibinfo
  {journal} {Optics Express}\ }\textbf {\bibinfo {volume} {26}},\ \bibinfo
  {pages} {24190} (\bibinfo {year} {2018})}\BibitemShut {NoStop}%
\bibitem [{\citenamefont {de~Aguiar}\ \emph {et~al.}(2016)\citenamefont
  {de~Aguiar}, \citenamefont {Gigan},\ and\ \citenamefont
  {Brasselet}}]{de_aguiar_enhanced_2016}%
  \BibitemOpen
  \bibfield  {author} {\bibinfo {author} {\bibfnamefont {H.~B.}\ \bibnamefont
  {de~Aguiar}}, \bibinfo {author} {\bibfnamefont {S.}~\bibnamefont {Gigan}}, \
  and\ \bibinfo {author} {\bibfnamefont {S.}~\bibnamefont {Brasselet}},\ }\href
  {\doibase 10.1103/PhysRevA.94.043830} {\bibfield  {journal} {\bibinfo
  {journal} {Physical Review A}\ }\textbf {\bibinfo {volume} {94}},\ \bibinfo
  {pages} {043830} (\bibinfo {year} {2016})}\BibitemShut {NoStop}%
\bibitem [{\citenamefont {Papadopoulos}\ \emph {et~al.}(2013)\citenamefont
  {Papadopoulos}, \citenamefont {Farahi}, \citenamefont {Moser},\ and\
  \citenamefont {Psaltis}}]{papadopoulos_high-resolution_2013}%
  \BibitemOpen
  \bibfield  {author} {\bibinfo {author} {\bibfnamefont {I.~N.}\ \bibnamefont
  {Papadopoulos}}, \bibinfo {author} {\bibfnamefont {S.}~\bibnamefont
  {Farahi}}, \bibinfo {author} {\bibfnamefont {C.}~\bibnamefont {Moser}}, \
  and\ \bibinfo {author} {\bibfnamefont {D.}~\bibnamefont {Psaltis}},\ }\href
  {\doibase 10.1364/BOE.4.000260} {\bibfield  {journal} {\bibinfo  {journal}
  {Biomedical Optics Express}\ }\textbf {\bibinfo {volume} {4}},\ \bibinfo
  {pages} {260} (\bibinfo {year} {2013})}\BibitemShut {NoStop}%
\bibitem [{\citenamefont {Caravaca-Aguirre}\ and\ \citenamefont
  {Piestun}(2017)}]{caravaca-aguirre_single_2017}%
  \BibitemOpen
  \bibfield  {author} {\bibinfo {author} {\bibfnamefont {A.~M.}\ \bibnamefont
  {Caravaca-Aguirre}}\ and\ \bibinfo {author} {\bibfnamefont {R.}~\bibnamefont
  {Piestun}},\ }\href {\doibase 10.1364/OE.25.001656} {\bibfield  {journal}
  {\bibinfo  {journal} {Optics Express}\ }\textbf {\bibinfo {volume} {25}},\
  \bibinfo {pages} {1656} (\bibinfo {year} {2017})}\BibitemShut {NoStop}%
\bibitem [{\citenamefont {Turtaev}\ \emph {et~al.}(2018)\citenamefont
  {Turtaev}, \citenamefont {Leite}, \citenamefont {Altwegg-Boussac},
  \citenamefont {Pakan}, \citenamefont {Rochefort},\ and\ \citenamefont
  {Čižmár}}]{turtaev_high-fidelity_2018}%
  \BibitemOpen
  \bibfield  {author} {\bibinfo {author} {\bibfnamefont {S.}~\bibnamefont
  {Turtaev}}, \bibinfo {author} {\bibfnamefont {I.~T.}\ \bibnamefont {Leite}},
  \bibinfo {author} {\bibfnamefont {T.}~\bibnamefont {Altwegg-Boussac}},
  \bibinfo {author} {\bibfnamefont {J.~M.~P.}\ \bibnamefont {Pakan}}, \bibinfo
  {author} {\bibfnamefont {N.~L.}\ \bibnamefont {Rochefort}}, \ and\ \bibinfo
  {author} {\bibfnamefont {T.}~\bibnamefont {Čižmár}},\ }\href {\doibase
  10.1038/s41377-018-0094-x} {\bibfield  {journal} {\bibinfo  {journal} {Light:
  Science \& Applications}\ }\textbf {\bibinfo {volume} {7}},\ \bibinfo {pages}
  {92} (\bibinfo {year} {2018})}\BibitemShut {NoStop}%
\bibitem [{\citenamefont {Badon}\ \emph {et~al.}(2016)\citenamefont {Badon},
  \citenamefont {Li}, \citenamefont {Lerosey}, \citenamefont {Boccara},
  \citenamefont {Fink},\ and\ \citenamefont {Aubry}}]{badon_smart_2016}%
  \BibitemOpen
  \bibfield  {author} {\bibinfo {author} {\bibfnamefont {A.}~\bibnamefont
  {Badon}}, \bibinfo {author} {\bibfnamefont {D.}~\bibnamefont {Li}}, \bibinfo
  {author} {\bibfnamefont {G.}~\bibnamefont {Lerosey}}, \bibinfo {author}
  {\bibfnamefont {A.~C.}\ \bibnamefont {Boccara}}, \bibinfo {author}
  {\bibfnamefont {M.}~\bibnamefont {Fink}}, \ and\ \bibinfo {author}
  {\bibfnamefont {A.}~\bibnamefont {Aubry}},\ }\href {\doibase
  10.1126/sciadv.1600370} {\bibfield  {journal} {\bibinfo  {journal} {Science
  Advances}\ }\textbf {\bibinfo {volume} {2}},\ \bibinfo {pages} {e1600370}
  (\bibinfo {year} {2016})}\BibitemShut {NoStop}%
\bibitem [{\citenamefont {Jeong}\ \emph {et~al.}(2018)\citenamefont {Jeong},
  \citenamefont {Lee}, \citenamefont {Choi}, \citenamefont {Kang},
  \citenamefont {Hong}, \citenamefont {Park}, \citenamefont {Lim},
  \citenamefont {Park},\ and\ \citenamefont {Choi}}]{jeong_focusing_2018}%
  \BibitemOpen
  \bibfield  {author} {\bibinfo {author} {\bibfnamefont {S.}~\bibnamefont
  {Jeong}}, \bibinfo {author} {\bibfnamefont {Y.-R.}\ \bibnamefont {Lee}},
  \bibinfo {author} {\bibfnamefont {W.}~\bibnamefont {Choi}}, \bibinfo {author}
  {\bibfnamefont {S.}~\bibnamefont {Kang}}, \bibinfo {author} {\bibfnamefont
  {J.~H.}\ \bibnamefont {Hong}}, \bibinfo {author} {\bibfnamefont {J.-S.}\
  \bibnamefont {Park}}, \bibinfo {author} {\bibfnamefont {Y.-S.}\ \bibnamefont
  {Lim}}, \bibinfo {author} {\bibfnamefont {H.-G.}\ \bibnamefont {Park}}, \
  and\ \bibinfo {author} {\bibfnamefont {W.}~\bibnamefont {Choi}},\ }\href
  {\doibase 10.1038/s41566-018-0120-9} {\bibfield  {journal} {\bibinfo
  {journal} {Nature Photonics}\ }\textbf {\bibinfo {volume} {12}},\ \bibinfo
  {pages} {277} (\bibinfo {year} {2018})}\BibitemShut {NoStop}%
\bibitem [{\citenamefont {Jackson}(2012)}]{jackson_towards_2012}%
  \BibitemOpen
  \bibfield  {author} {\bibinfo {author} {\bibfnamefont {S.~D.}\ \bibnamefont
  {Jackson}},\ }\href {\doibase 10.1038/nphoton.2012.149} {\bibfield  {journal}
  {\bibinfo  {journal} {Nature Photonics}\ }\textbf {\bibinfo {volume} {6}},\
  \bibinfo {pages} {423} (\bibinfo {year} {2012})}\BibitemShut {NoStop}%
\end{thebibliography}%


\begin{thebibliography}{1}

\bibitem{carpenter_observation2015} J. Carpenter et al., {\it Nat. Photon.} {\bf 11} 751 (2015).
\bibitem{gerchberg_practical_1972} R. W Gerchberg et al., {\it Optik} {\bf 35} 237 (1972).
\end{thebibliography}

\pagebreak

\begin{center}
\textbf{\large Control of the temporal and polarization response of a multimode fiber: supplementary material}
\end{center}
\setcounter{equation}{0}
\setcounter{figure}{0}
\setcounter{table}{0}
\setcounter{section}{0}
\makeatletter
\renewcommand{\theequation}{S\arabic{equation}}
\renewcommand{\thefigure}{S\arabic{figure}}
\renewcommand{\thesection}{S\arabic{section}}
\renewcommand{\bibnumfmt}[1]{[S#1]}
\renewcommand{\citenumfont}[1]{S#1}

\section{Additional information on the experimental measurement of the Multi Spectral Transmission Matrix}

The experimental setup presented in Figure 1 enables us to shape the input Gaussian beam from the laser into an arbitrary spatial and polarization state. A polarization controller (not shown in Figure 1) aligns the input polarization state such that the power is split equally between the H and V ports of the first polarization beam splitter. The vertical polarization beam is rotated with an half waveplate, as required for the LCOS-SLM beam shaping. These two beams are then independently manipulated on the left and the right sides of the spatial light modulator~\cite{carpenter_observation2015}. For clarity purposes, on Figure 1 the SLM is illustrated as transmittive, however in reality the SLM is used in reflection. The two polarization beams are then recombined using polarization optics and imaged onto the input facet of the multimode fiber.

In order to achieve amplitude and phase modulation of the input beam with a phase-only SLM, spatial filtering in the Fourier plane of the SLM is required. This spatial filtering is performed by the multimode fiber itself, which is located in the first diffractive order of the SLM in the Fourier plane. 

The transmission matrix is measured in the mode basis of the multimode fiber. The $N_{\text{input}}/2$ modes per polarization are sent one after another onto the input facet of the multimode fiber. For a given input mode (for either the horizontal or the vertical polarization), the phase pattern to be displayed onto the SLM is calculated with the Gerchberg-Saxton algorithm~\cite{gerchberg_practical_1972}. After propagation through the multimode fiber, the transmitted output field is then measured with the digital off-axis holography setup simultaneously for the two polarization states. The output field is overlapped with the $N_{\text{output}}/2$ modes per polarization. We decompose the output field in the LG basis for computation complexity reasons. The decomposition is made with many more modes until almost all power is accounted for, which explains why we have $N_{\text{output}}~>~N_{\text{input}}$. The mode decomposition at the output, for a specific input mode, forms a column of the transmission matrix. 

\begin{figure*}[htbp]
\centering
\includegraphics[width=\linewidth]{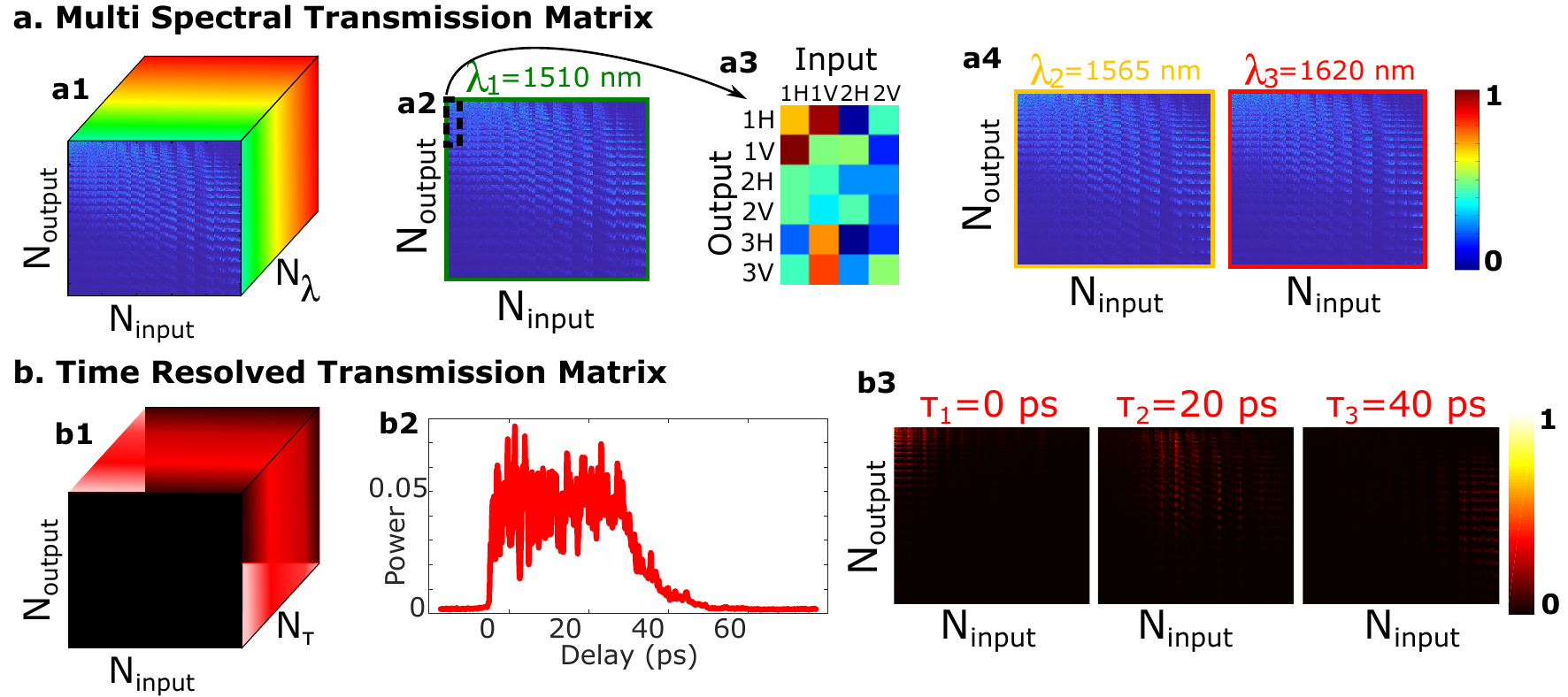}
\caption{\textbf{(a)} Multi Spectral Transmission matrix of the multimode fiber. (a1) Artistic representation of the experimentally measured MSTM. Its dimensions are $N_{\text{input}}=254$~input modes, $N_{\text{output}}=650$~output modes, and $N_{\lambda}=1273$~ spectral components. (a2) Amplitude of the transmission matrix measured at $\lambda_1=1510$~nm. (a3) Zoom of (a2). The polarization information is interlaced both at the input and at the output. $ij$ stands for mode $i$, $i$ is an integer from 1 to $N_{\text{input}}/2$, on the polarization state $j$, $j$ being either H (for horizontal) or V (for vertical). (a4) Amplitude of the transmission matrices measured at $\lambda_2=1565$~nm and $\lambda_3=1610$~nm. \textbf{(b)} Time Resolved Transmission matrix of the multimode fiber. (b1) Artistic representation of the TRTM, calculated with a Fourier transform of the experimentally measured MSTM along the spectral axis. (b2) Total power of the transmission matrix per delay time. (b3) Amplitude of the transmission matrices at different delay times: $\tau_1=0$~ps, $\tau_2=20$~ps and $\tau_3=40$~ps. }
\label{fig:MSTM_TRTM}
\end{figure*}

The Multi Spectral Transmission Matrix $U(\lambda)$ is the stack of transmission matrices for all the measured spectral components. Figure \ref{fig:MSTM_TRTM}a shows the amplitude of three transmission matrices measured at different wavelengths. Figure \ref{fig:MSTM_TRTM}a3 shows a zoom of the amplitude of  transmission matrix at $\lambda=1510$~nm. For a given input mode, the polarization content of the output field is interlaced, which defers slightly with the way $U(\lambda)$ is written in Equation 4. Nonetheless, one can re-organize the mode decomposition coefficients within the transmission matrix to match Equation 4.

\section{Correlation between Time Resolved transmission matrices}

The Time Resolved Transmission matrix is calculated with the experimentally measured MSTM, with a Fourier transform along the spectral axis. Figure \ref{fig:MSTM_TRTM}b2 illustrates the total power per delay. Figure \ref{fig:MSTM_TRTM}b3 shows the transmission matrix at three different delay times: 0~ps, 20~ps and 40~ps. The low order modes tend to have most of their power at early arrival times, while the higher order modes have most of their power at longer arrival times. 

In a step-index multimode fiber, the impulse response of each mode is broad in the time domain, as each input mode is coupled along propagation with other modes that have different group delays. When we manipulate the impulse response at different delay times, the impulse response could interfere if the delay times are close to each other. We can quantify this minimum time interval where the transmission matrices won't be correlated to each other with the temporal correlation C($\tau$,$\tau'$). We define the correlation C($\tau$,$\tau'$) as the overlap between the transmission matrix U($\tau$) at delay $\tau$ and the transmission matrix U($\tau'$) at delay $\tau'$:

\begin{equation}
C(\tau,\tau')= \Bigg\vert \dfrac{\sum_{i=1}^{N_{\text{input}}} \sum_{j=1}^{N_{\text{output}}} U_{ij}(\tau) U^{*}_{ij}(\tau')}{\sqrt{\sum_{i} \sum_{j} \vert U_{ij}(\tau) \vert^2} \sqrt{\sum_{i} \sum_{j} \vert U_{ij}(\tau') \vert^2} }\Bigg\vert   
\end{equation}
with $U_{ij}(\tau)$ the coefficient of the transmission matrix at delay $\tau$ relating the $i-th$ input mode and the $j-th$ output mode, and $^*$ the complex conjugate operator. With this definition, we thus have C($\tau$,$\tau$)=1, which corresponds to the auto-correlation of the time resolved transmission matrix at time $\tau$.

\begin{figure}[b!]
\centering
\includegraphics[width=\linewidth]{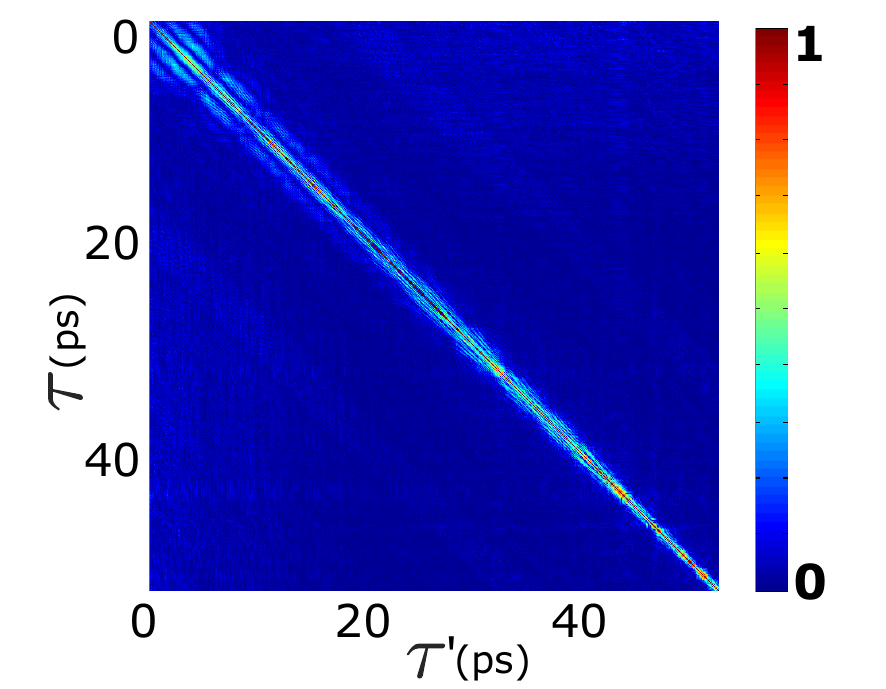}
\caption{2D representation of the temporal correlation C($\tau$,$\tau'$) between time resolved transmission matrices at delay $\tau$ and $\tau'$, for a set of $\tau$ and $\tau'$ varying between 0~ps and 50~ps.}
\label{fig:correlation_TRTM}
\end{figure}

Figure~\ref{fig:correlation_TRTM} shows the 2D map of C($\tau$,$\tau'$) for $\tau$ and $\tau'$ between 0~ps and 50~ps. On average, for a given delay $\tau$, the correlation C($\tau$,$\tau + \delta\tau$) is negligible (e.g  C($\tau$,$\tau + \delta\tau$)<0.1 ) for $\delta\tau \sim 1-2$~ps.

\section{Multiple delay control: additional examples}

\begin{figure*}[htbp]
\centering
\includegraphics[width=\linewidth]{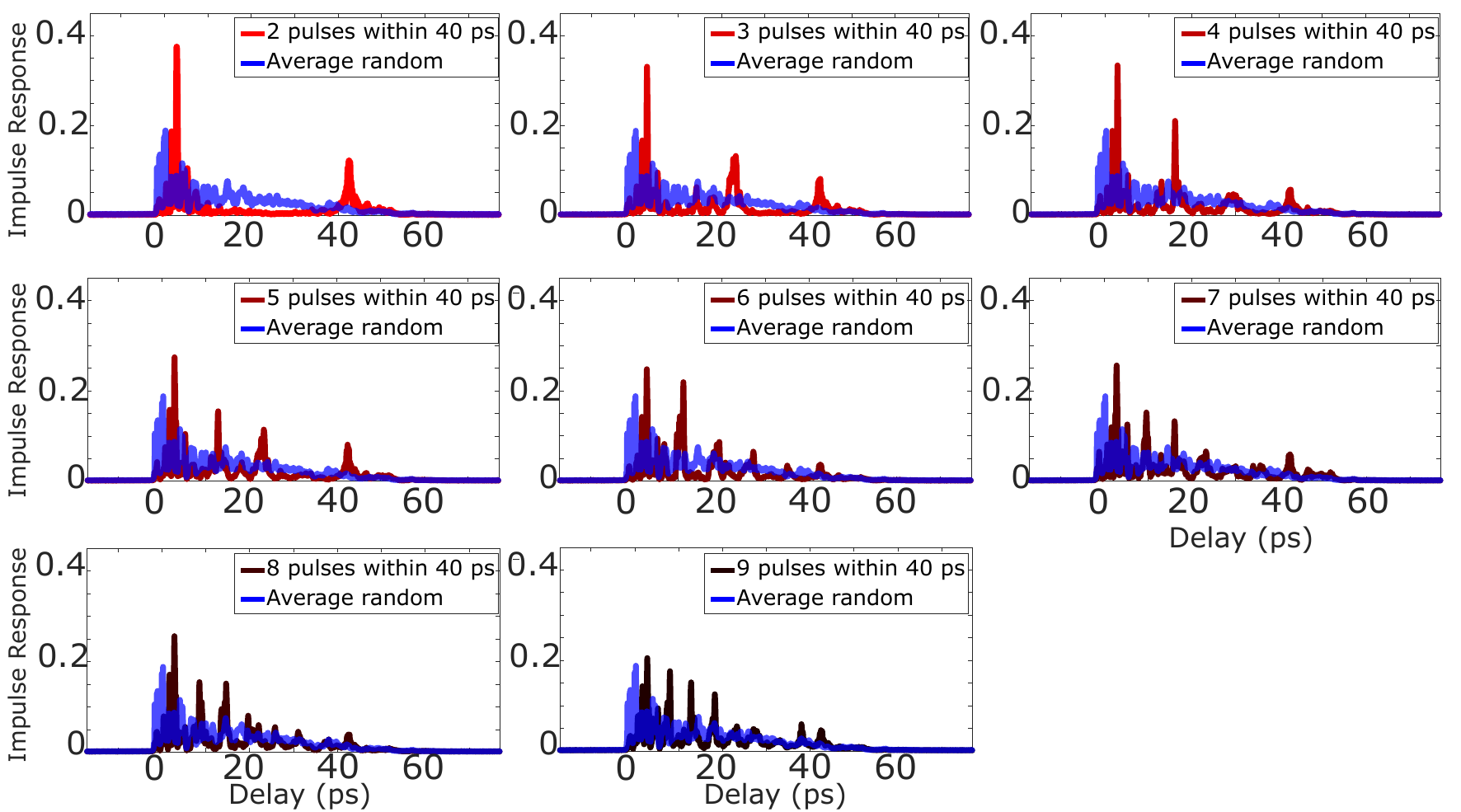}
\caption{Enhancement of transmitted light intensity at $N_{\text{superpos.}}$ multiple delay times between 4~ps and 44~ps, $N_{\text{superpos.}}$ varying from 2 to 9.}
\label{fig:multiple_40ps}
\end{figure*}

\begin{figure*}[htbp]
\centering
\includegraphics[width=\linewidth]{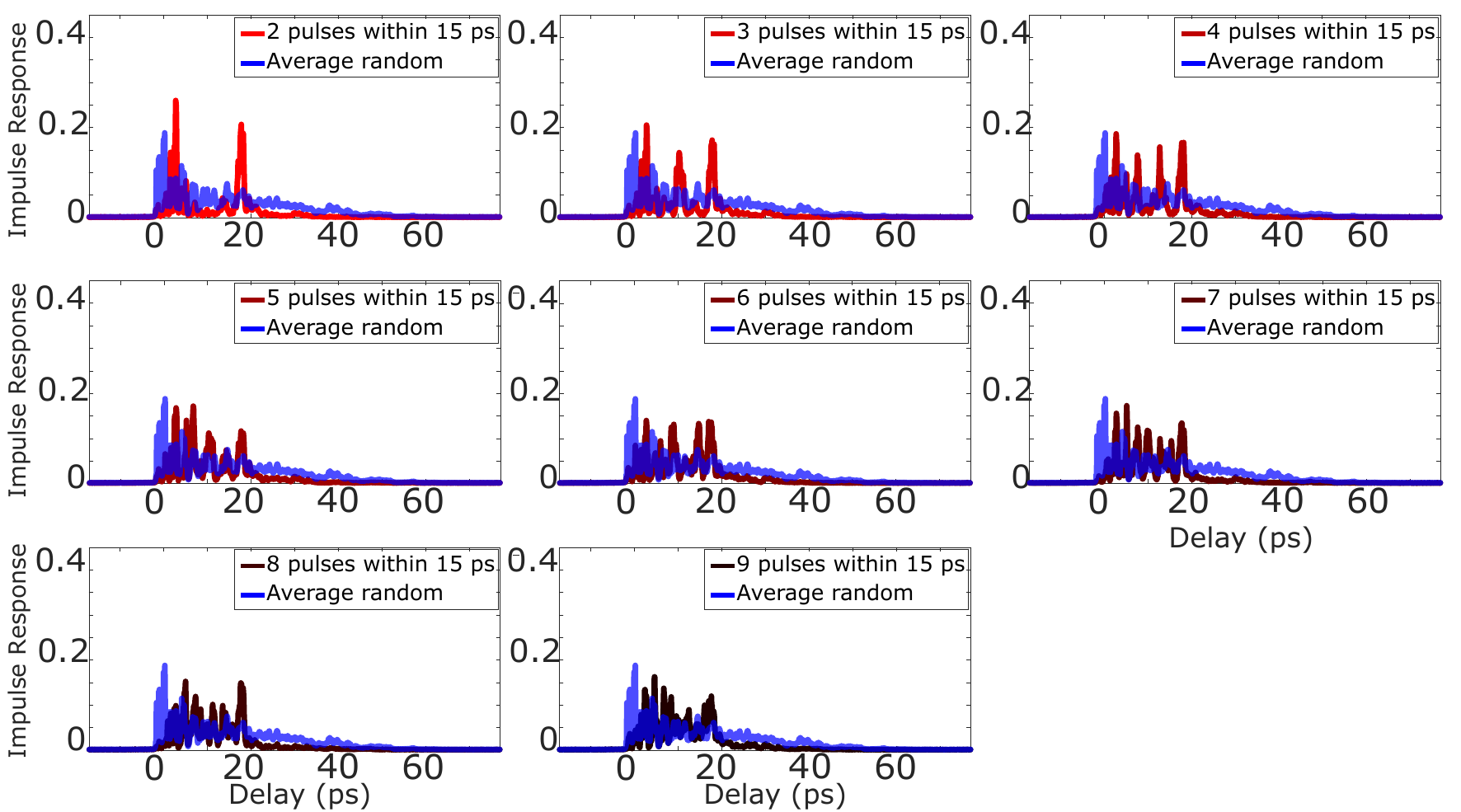}
\caption{Enhancement of transmitted light intensity at $N_{\text{superpos.}}$ multiple delay times between 4~ps and 19~ps, $N_{\text{superpos.}}$ varying from 2 to 9.}
\label{fig:multiple_15ps}
\end{figure*}

In Figure 5, we show multiple delay control by superposing $N_{\text{superpos.}}$ eigenstates at different arrival times. More precisely, we show a superposition of $N_{\text{superpos.}}=6$~solutions within a 40~ps time interval, and $N_{\text{superpos.}}=7$~solutions within a 15~ps time interval. We provide additional superposition plots in Figure~\ref{fig:multiple_40ps} and Figure~\ref{fig:multiple_15ps}, with $N_{\text{superpos.}}$ varying from 2 to 9, within 40~ps in Figure~\ref{fig:multiple_40ps} and within 15~ps in Figure~\ref{fig:multiple_15ps}.

\end{document}